\documentclass[]{pasj00}

\begin{document}
\SetRunningHead{Asai et.al.}{Outbursts of Aql~X-1 and 4U~1608$-$52}
\Received{2012/02/25}
\Accepted{2012/06/18}
\Published{2013/02/25}

\title{Slow and Fast Transitions in the Rising Phase of Outbursts 
from NS-LMXB transients, Aql~X-1 and 4U~1608$-$52}

\author{
Kazumi \textsc{Asai},\altaffilmark{1}
Masaru \textsc{Matsuoka},\altaffilmark{1}
Tatehiro \textsc{Mihara},\altaffilmark{1}
Mutsumi \textsc{Sugizaki},\altaffilmark{1}
Motoko \textsc{Serino},\altaffilmark{1}
Satoshi \textsc{Nakahira},\altaffilmark{1}
Hitoshi \textsc{Negoro},\altaffilmark{2}
Yoshihiro \textsc{Ueda},\altaffilmark{3}
and Kazutaka\textsc{Yamaoka}\altaffilmark{4}
}
\altaffiltext{1}{MAXI team, RIKEN, 2-1 Hirosawa, Wako, Saitama 351-0198}
\email{kazumi@crab.riken.jp}
\altaffiltext{2}{Department of Physics, Nihon University,
1-8-14 Kanda-Surugadai, Chiyoda-ku, Tokyo 101-8308}
\altaffiltext{3}{Department of Astronomy, Kyoto University,
Kitashirakawa, Oiwake-cho, Sakyo-ku, Kyoto 606-8502}
\altaffiltext{4}{Department of Physics and Mathematics,
Aoyama Gakuin University, 5-10-1 Fuchinobe, Chuo-ku, Sagamihara 252-5258}

\maketitle

\begin{abstract}

We analyzed the initial rising behaviors of X-ray outbursts from two transient
low-mass X-ray binaries (LMXBs) containing a neutron-star (NS),
Aql~X-1 and 4U~1608$-$52, which are
continuously being monitored by MAXI/GSC in 2--20~keV, RXTE/ASM in 2--10~keV,
and Swift/BAT in 15--50~keV.
We found that the observed ten outbursts are classified into two types
by the patterns of the relative intensity evolutions in the two energy bands
below/above 15~keV.
One type behaves as the 15--50~keV intensity achieves the maximum during
the initial hard-state period and drops greatly at  the hard-to-soft state
transition.
On the other hand, the other type does as both the 2--15~keV and the 15--50~keV
intensities achieve the maximums after the transition.
The former have the longer initial hard-state
($\gtrsim$ 9 d) than the latter's
($\ltsim$5 d).
Therefore, we named them as slow-type  (S-type) and fast-type (F-type),
respectively.
These two types also show the differences in the luminosity at the hard-to-soft
state transition as well as in the average luminosity before
the outburst started, where the S-type are higher than the F-type in the both.
These results suggest that the X-ray radiation during the pre-outburst period,
which heats up the accretion disk and delays the disk transition
(i.e., from a geometrically thick disk to a thin one),
would determine whether the following outburst becomes S-type or F-type.
The luminosity when the hard-to-soft state transition occurs is
higher than $\sim 8 \times10^{36}$ erg s$^{-1}$ in the S-type,
which corresponds to 4\% of the Eddington luminosity for a 1.4 \Mo NS.
\end{abstract}

\section{Introduction}
Soft X-ray transients (SXTs) are a group of X-ray binary systems
that occasionally exhibit bright outbursts in a soft X-ray band.
They are mostly identified as a low-mass X-ray binary (LMXB)
containing either a neutron-star (NS-LMXB) or a black hole (BH-LMXB).
The luminosity of the NS-LMXBs increases by
2--5 orders of magnitude
during the outbursts.
As the luminosity increases, the spectral state usually changes from the
low/hard state to the high/soft state in both
the NS-LMXB and the BH-LMXB
(e.g., \cite{Klis1994}; \cite{Campana1998}; \cite{Remillard2006}).
The low/hard state has a low intensity and a very hard spectrum, 
where the Comptonized component is dominant.
On the other hand, the high/soft state has a high intensity and
a very soft spectrum, where the thermal component is dominant.
We study the initial rising phases of the NS-LMXB outbursts
in this paper.

NS-LMXBs have been classified into two groups,
Z sources and Atoll sources, based
on the behavior on the color--color diagram (\cite{Hasinger1989}).
Z sources are generally bright and sometime become close to
the Eddington luminosity limit.
On the other hand, Atoll sources are generally less bright, and
some of them show transient activity.
The luminosity changes by three order of magnitudes
from the quiescent state to the active state,
whose luminosity also reaches close to the Eddington limit at the maximum.
\citet{Homan2010} suggested that the differences between Z and Atoll sources
are originated from the mass accretion rate.

The hard-to-soft state transition
in the Atoll sources is also mainly due to the mass accretion rate.
However in some transient objects, the mass accretion rate
is not the only parameter that determines the
\textcolor{black}{hard-to-soft state transition}
(\cite{Yu2007}; Yu et al. 2004; \cite{Smith2002}; \cite{Homan2001}).
Therefore, it has not been completely understood which parameter actually
derives the \textcolor{black}{hard-to-soft state transition}.

The hard-to-soft transition in the rising phase and
the soft-to-hard transition in the decay phase
have been studied separately.
Yu and Yan (2009) and Tang, Yu, and Yan (2011) analyzed the relation of
the X-ray luminosity at the hard-to-soft transition to the peak
luminosity during the subsequent soft state as well as
to the luminosity-increasing rate at the transition in both
the NS-LMXBs and the BH-LMXBs of both the
transients and the persistents. 
They concluded from the results that the hysteresis between
the hard-to-soft and the soft-to-hard state transitions is primarily
due to a non-stationary accretion flow when mass-accretion rate increases.
\citet{Maccarone2003} reported that the soft-to-hard transition in
the decay phase occurs at a luminosity of 1\%--4\% of the Eddington limit.
Maccarone and Coppi (2003) suggested that the propeller mechanism
is not the sole cause for the state transition in NS-LMXBs from
the results of RXTE observations of Aql~X-1, which showed that
the luminosity in the hard-to-soft transition in the rising
phase is greater than that in the soft-to-hard transition by
a factor of $\sim$ 5 or more.
Such a hysteric behavior was first pointed out by \citet{Miyamoto1995}
for the rising phase of BH-LMXBs.

X-ray spectra of NS-LMXBs basically show a monotonous continuum.
However, the emission model is still controversial because
they exhibit complex time variations linked with the change of
the spectral states.
A variety of spectral models have been proposed so far to explain
their entire behaviors with a unified scheme.
They are represented by two major models,
the Eastern model (after \cite{Mitsuda1984}, 1989)
and the Western model (after \cite{White1988}).
These models generally consist of two components,
a blackbody radiation with one or more temperatures and
a Comptonized radiation extending to the hard X-ray band 
(\cite{Barret2001}; \cite{Gierlinski2002}; \cite{Lin2007};
\cite{Farinelli2008}; \cite{Takahashi2011}; \cite{Sakurai2012}).
In the high/soft state, the blackbody radiation is dominant where
the Comptonized radiation still exists in the low level 
(\cite{Gierlinski2002}; \cite{Paizis2006}; \cite{Lin2007}; \cite{Raichur2011}).
The Comptonized component is pointed out to be a major component
in the low/hard state 
(
\textcolor{black}{
e.g., Mitsuda et al. 1989}
).
The site of the high-energy electrons responsible
for the Comptonization and the seed photons are still a source of arguments.

4U~1608$-$52 and Aql~X-1 are famous transient NS-LMXBs,
\textcolor{black}{
where Type I X-ray bursts have also been detected in both sources}
(4U~1608$-$52: \cite{Nakamura1989}, Aql~X-1: \cite{Koyama1981}).
Their outbursts typically show a sharp rise and an exponential decay.
Campana et al. (1998) reported that the rise time is 5--10~d and
the 
\textcolor{black}{
exponential decay time
}
is 30--70~d in Aql~X-1, and also that
4U~1608$-$52 sometimes occurs outbursts with a symmetric evolution
between the rise and the decay. 
The periodicity of the outbursts has been reported in both 4U~1608$-$52
(e.g., \cite{Simon2004}) and Aql~X-1 (\cite{Kitamoto1993}; \cite{Simon2002}).
\citet{Chen2006} reported that 4U~1608$-$52 
occurs the soft-to-hard transition when the luminosity is in
(3.3--5.3)$\times10^{36}$~ergs s$^{-1}$,
assuming that the distance to the source is 3.6~kpc.
As for Aql~X-1, it is reported that the hard-to-soft transition occurs
at the luminosity of (4.2--5.5)$\times10^{36}$~ergs s$^{-1}$
and the soft-to-hard
transition at (6.1--7.5)$\times10^{35}$~ergs s$^{-1}$
assuming the distance of 2.5~kpc
(\cite{MacCop2003}; \cite{Maccarone2003}).

In this paper, we focus on the behaviors of the rising
phase of outbursts from 4U~1608$-$52 and Aql~X-1.
We investigate the evolution of their spectral states using the
X-ray light curves and the hardness ratio obtained from
the data of continuous monitoring with
MAXI/GSC in 2--20~keV, RXTE/ASM in 2--10~keV, and
Swift/BAT in 15--50~keV.
We classify all observed outbursts by the pattern of the state transition
during the rising phase.
We then analyze their differences and discuss what determines
the different evolution patterns.
\textcolor{black}{
We employ the distances of $4.1\pm0.4$~kpc for 4U~1608$-$52 and that of
$5.0\pm0.9$~kpc
for Aql~X-1 following \citet{Galloway2008} hereafter throughout the paper.
}

\section{Analysis and Results}

\subsection{MAXI-GSC, Swift-BAT Light Curves and the Hardness Evolution}
\label{MAXI/GSC and BAT section}

\begin{figure}
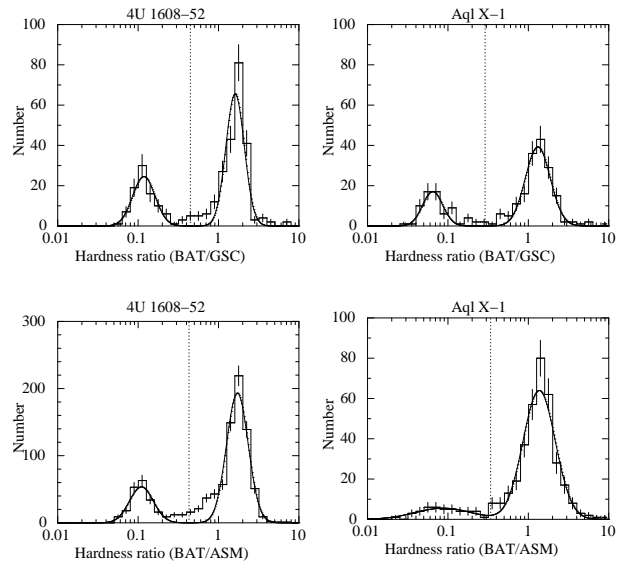

  \begin{center}
    \FigureFile(40mm,){fig1-1.eps}
    \FigureFile(40mm,){fig1-2.eps}
  \end{center}
  \begin{center}
    \FigureFile(40mm,){fig1-3.eps}
    \FigureFile(40mm,){fig1-4.eps}
  \end{center}
  \caption{Distributions of hardness ratios of BAT/GSC (top) and
BAT/ASM (bottom) on a one-day timescale for 4U~1608$-$52 (left)
and Aql~X-1 (right), 
\textcolor{black}{
where the vertical error bars represent 1-$\sigma$ statistical uncertainty.
}
The best-fit
\textcolor{black}{
model of two log-normal distributions
}
are overlaid.
The vertical dotted lines represent the thresholds between the soft and
the hard states which we adopted.
}
\label{fig1}
\end{figure}

MAXI (Monitor of All-sky X-ray Image: Matsuoka et al. 2009) on the
International Space Station (ISS)
started the all-sky survey
on 2009 August 15 with GSC
(Gas Slit Camera: \cite{Mihara2011}; \cite{Sugizaki2011a}) 
in 2--30~keV and SSC (Solid-state Slit Camera: Tomida et al. 2011)
in 0.5--12 keV.
The GSC with MAXI nova alert system (Negoro et al. 2010)
has detected two outbursts from
4U~1608$-$52 in 2010 March \citep{Morii2010} and in
2011 March \citep{Sugizaki2011b} and three outbursts from 
Aql~X-1 in 2009 November, 2010 September,
and 2011 October \citep{Yamaoka2011}. 
We obtained the GSC light curve data in 2--20~keV band from the public archive~
\footnote{$<$http://maxi.riken.jp/$>$.}
provided by MAXI team. 
We also obtained the light-curve data in the hard X-ray band of
15--50~keV monitored by BAT (Burst Alert Telescope: Barthelmy et al. 2005)
on board Swift (Gehrels et al. 2004) from the transient-monitor results
provided by the Swift-BAT team.
\footnote{$<$http://heasarc.gsfc.nasa.gov/docs/swift/results/transients/$>$.}

We identified the spectral state from the BAT/GSC hardness ratio
using the same method employed in
Yu and Yan (2009) and 
Tang, Yu, and Yan (2011).
X-ray spectra of both 4U~1608$-$52 and Aql~X-1 can be approximated by a
two-component model consisting of a
\textcolor{black}{
blackbody emission
}
and
a Comptonized emission (figure~2 in \cite{Gierlinski2002} for 4U~1608$-$52 and
figure~5 in \cite{Lin2007} for Aql~X-1).
In the soft state, the blackbody emission with a temperature
of kT $\simeq$ 0.5--2.0~keV becomes significant and dominates
the GSC 2--20~keV band.
Meanwhile, the flux in the BAT band of 15--50~keV are mostly dominated
by the Comptonized emission which often extends to several tens keV.
Therefore, the ratio represents the spectral state.

To determine the best threshold to discriminate the two states,
we investigated the distributions of the daily hardness ratio
of the BAT 15--50~keV flux to the 2--20~keV GSC flux 
as shown in figure~1. 
The fluxes are converted to that in Crab unit using the nominal Crab
values of 3.6~photons s$^{-1}$ cm$^{-2}$ in the GSC (cf. footnote 1)
and 0.22~counts s$^{-1}$ cm$^{-2}$ in the BAT
(cf. footnote 2).
We used all of the data points obtained from 2009 August 15 (MJD=55058) 
to 2011 December 19 (MJD=55914) after MAXI started the all-sky survey.
The distributions of both 4U~1608$-$52 and Aql~X-1 clearly show two peaks
corresponding to the soft and the hard states.
\textcolor{black}{
 We fitted the histogram with a 
model of two log-normal distributions and determined
 the soft-hard threshold at the center of the two peaks obtained
 from the best-fit parameters.
 The center values of the Gaussian ($C_1$ and $C_2$) are
 $\log_{10} C_1 = 0.21 \pm 0.02$, $\log_{10} C_2 = -0.93 \pm 0.03$ and
 $\log_{10} C_1 = 0.12 \pm 0.02$, $\log_{10} C_2 = -1.19 \pm 0.04$,
 for 4U~1608$-$52 and Aql~X-1, respectively.
 The obtained thresholds, $10^{(\log_{10} C_1+ \log_{10} C_2)/2}$, 
 are 0.44 and 0.29, respectively.
}

\begin{figure*}
  \begin{center}
    \FigureFile(80mm,){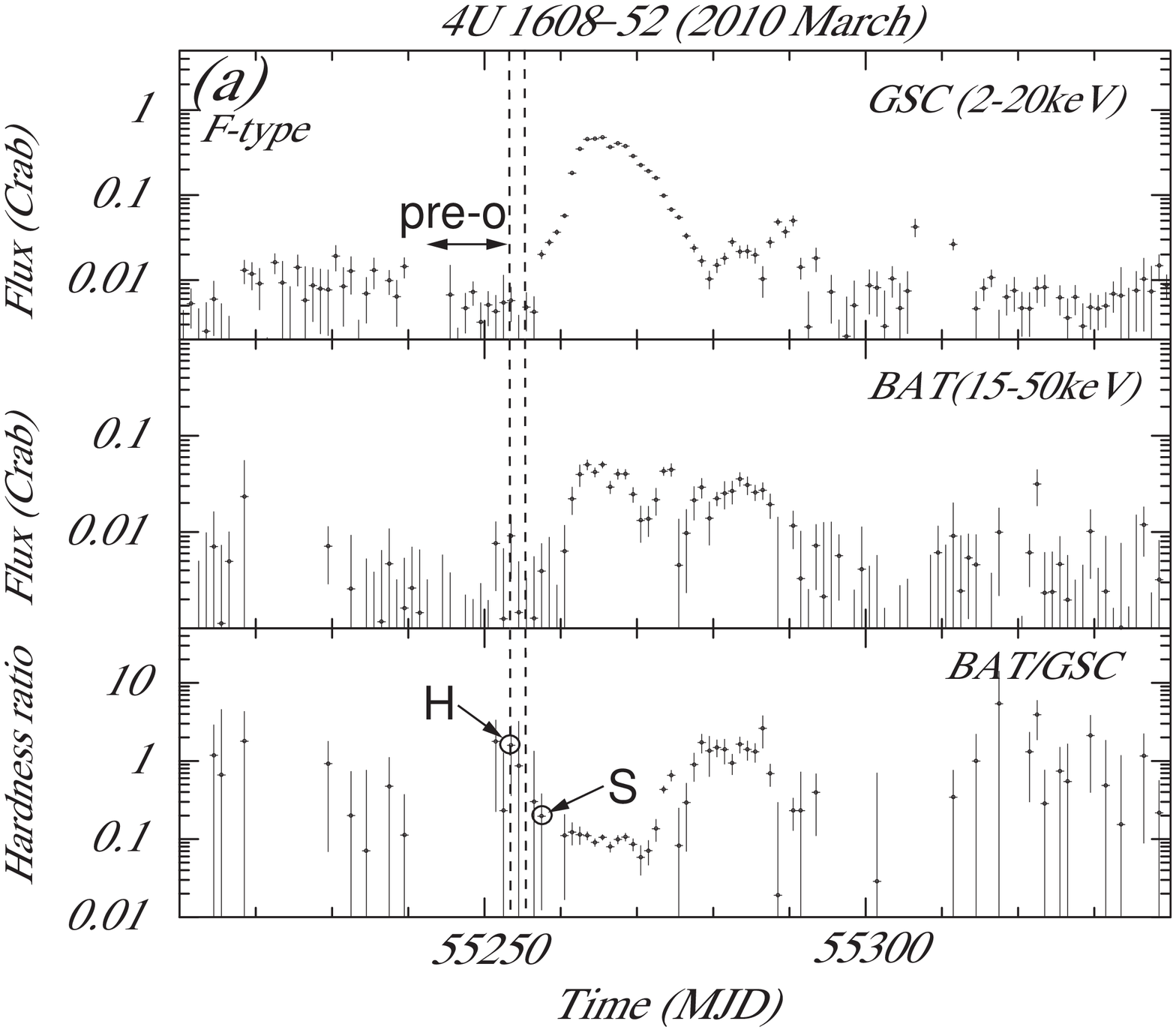}
    \FigureFile(80mm,){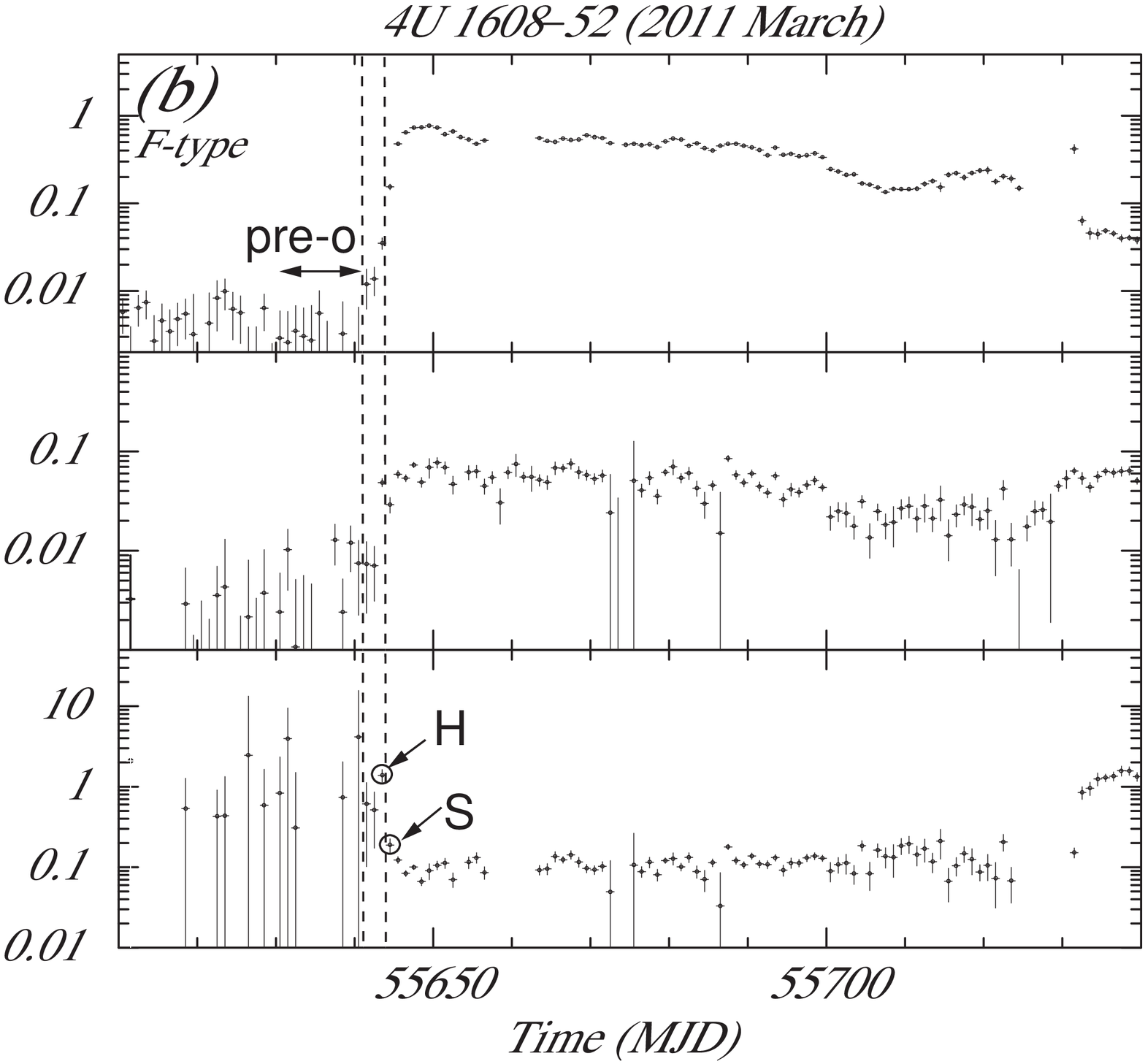}
    \FigureFile(80mm,){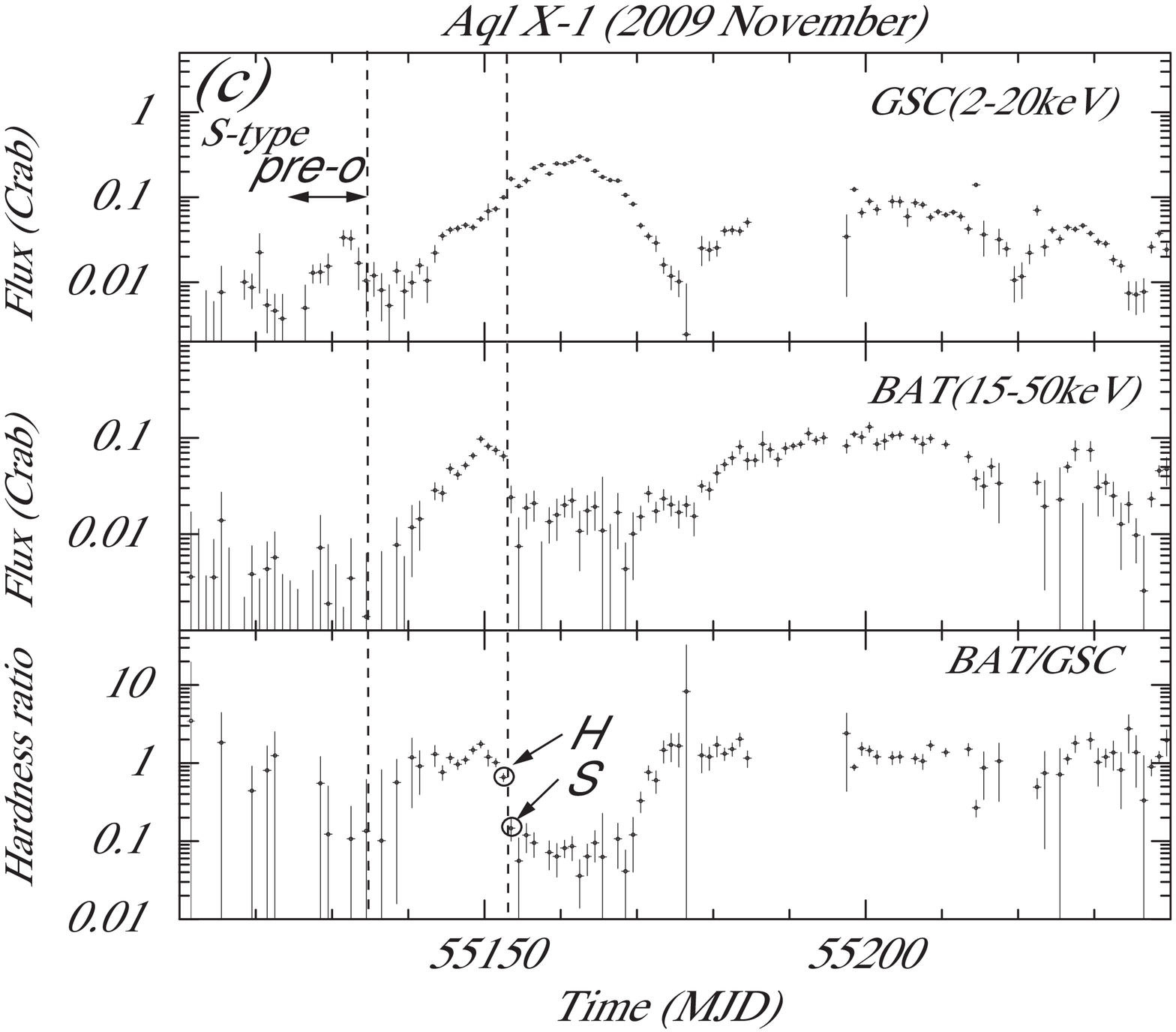}
    \FigureFile(80mm,){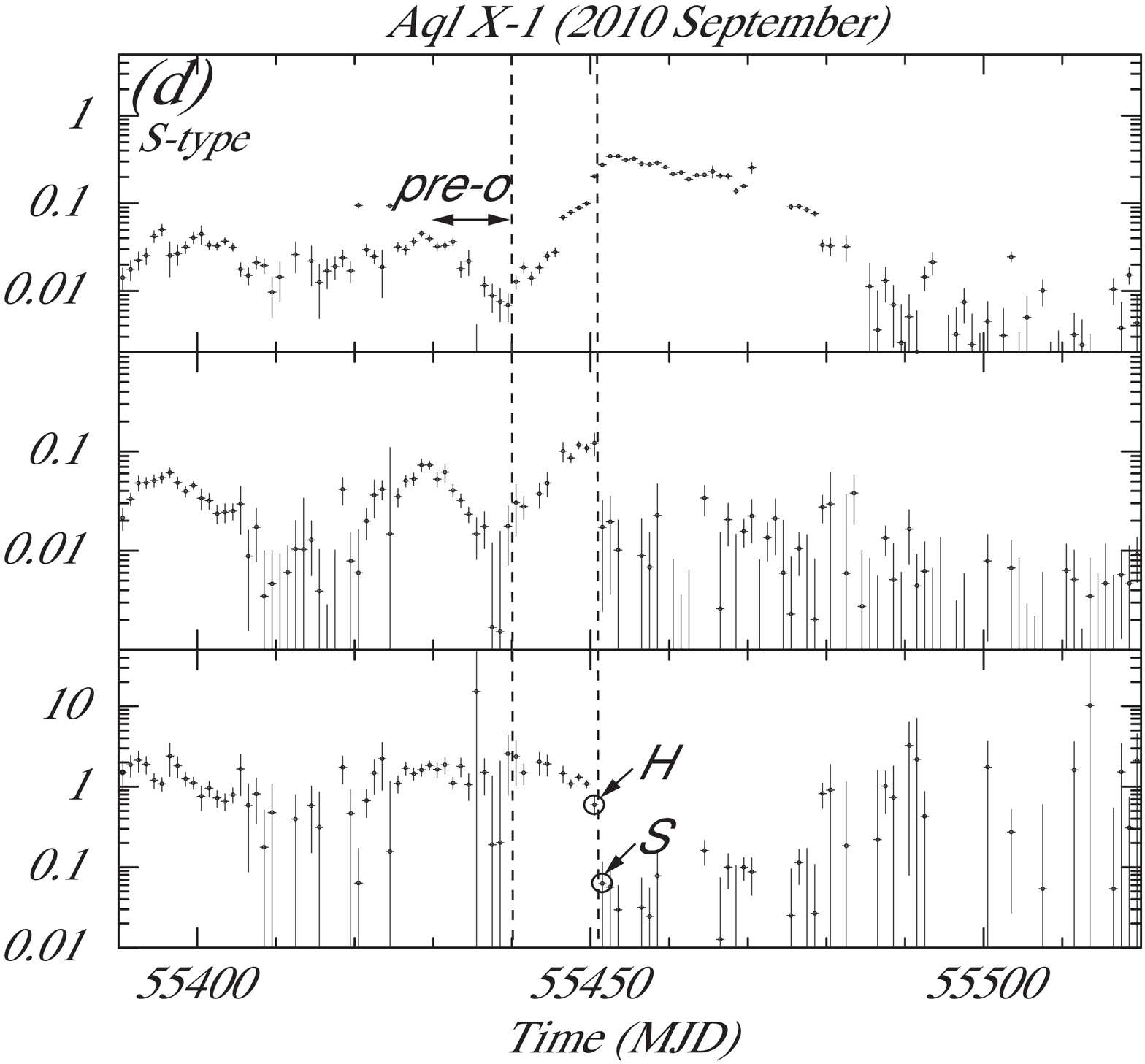}
    \FigureFile(80mm,){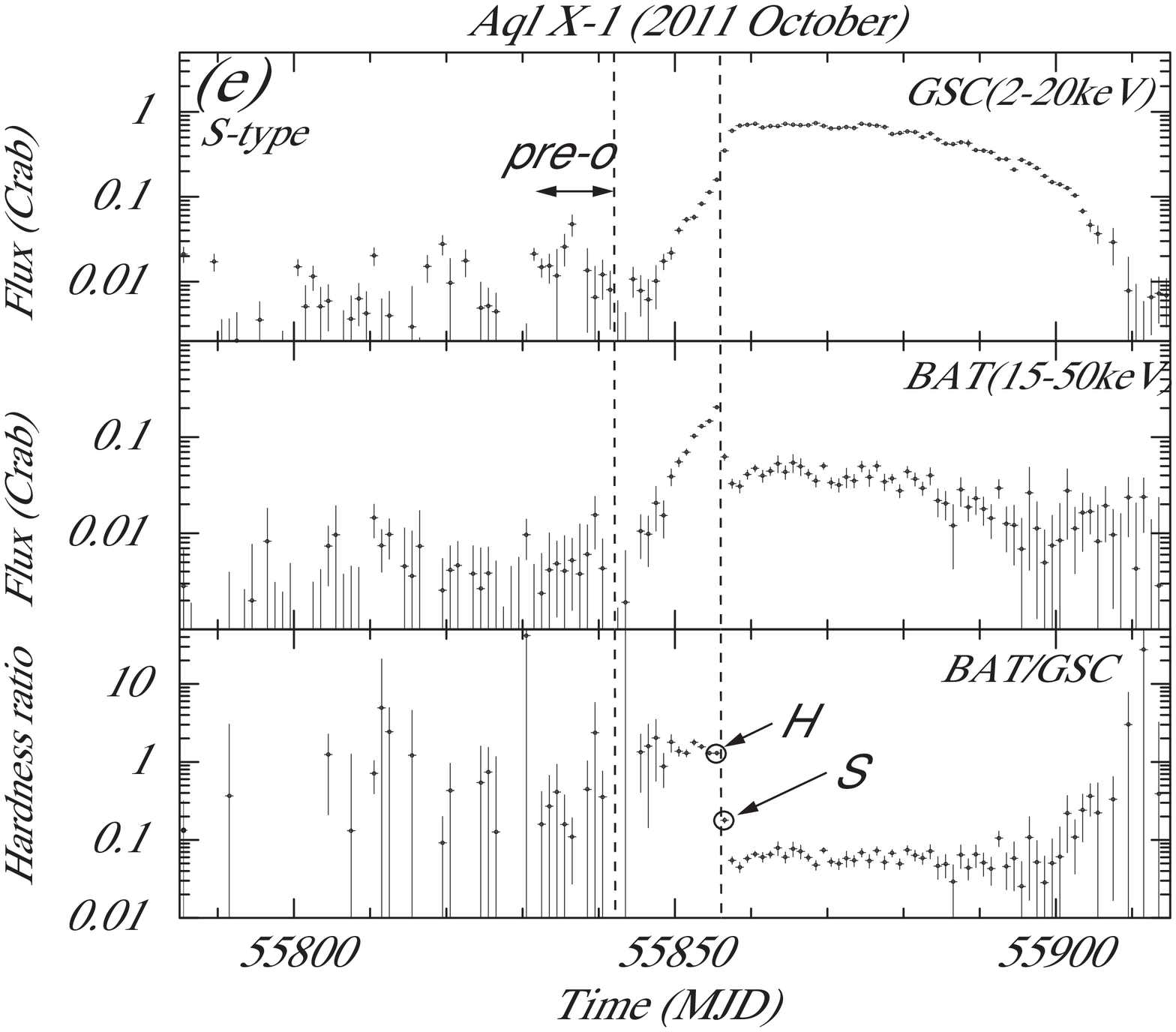}
  \end{center}
  \caption{MAXI-GSC light curves (2--20~keV),
Swift-BAT light curves (15--50~keV),
and the hardness ratios (BAT/GSC) of five outbursts.
The vertical error bars represent 1-$\sigma$ statistical uncertainty.
The labels ``H'' and ``S'' are defined
in subsection \ref{MAXI/GSC and BAT section} as
the last point of the hard state and
the first point of the soft state, respectively.

The two vertical dashed lines indicate the time at the outburst onset
(left) and that at the hard-to-soft transition (right) that we derived.
The ``pre-o'' denotes the pre-outburst period used
in subsection \ref{Luminosity before Outburst section}, which is 10~d
before the outburst onset.}
\label{fig2}
\end{figure*}

\begin{figure*}
  \begin{center}
    \FigureFile(70mm,){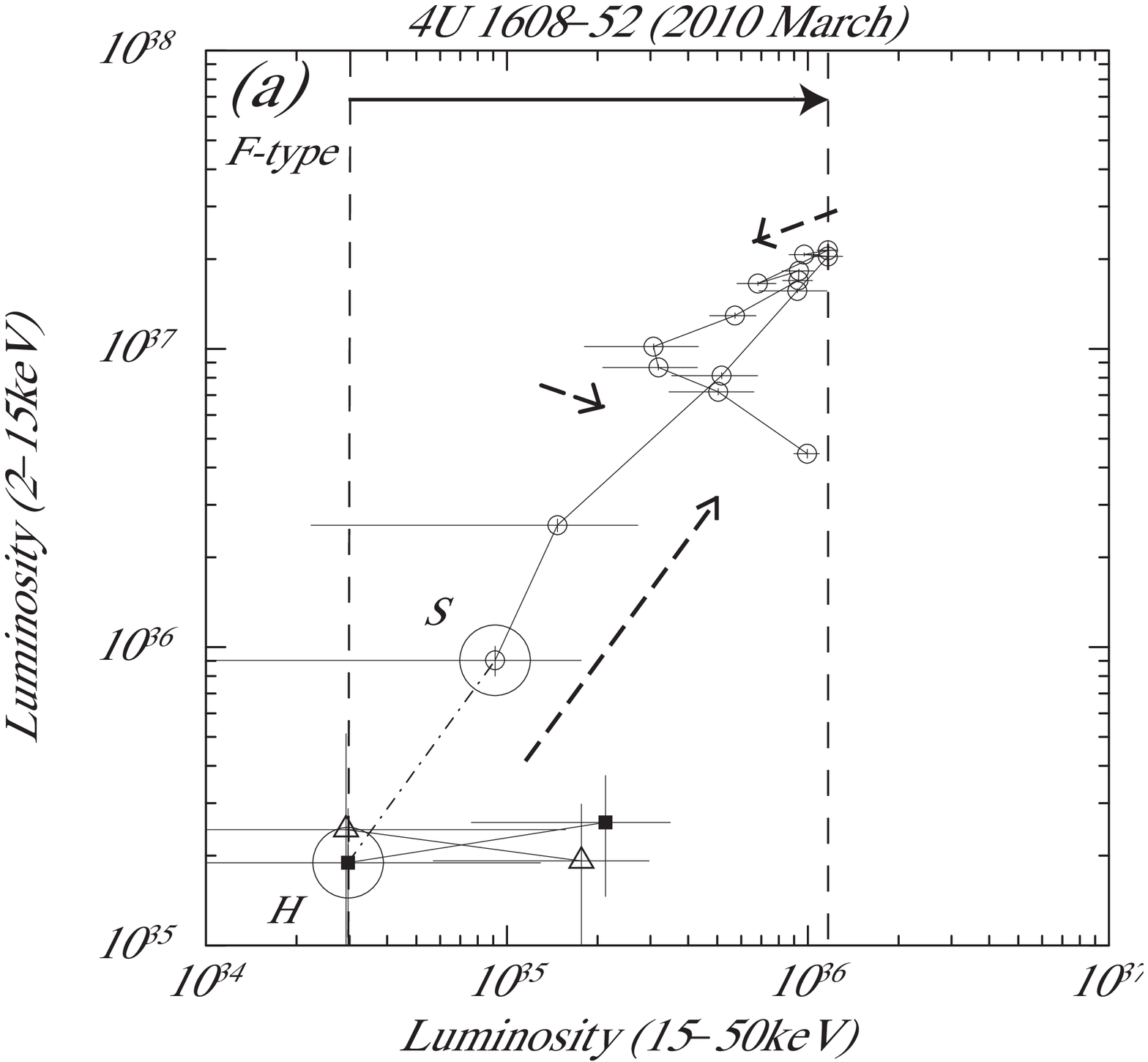}
    \FigureFile(64mm,){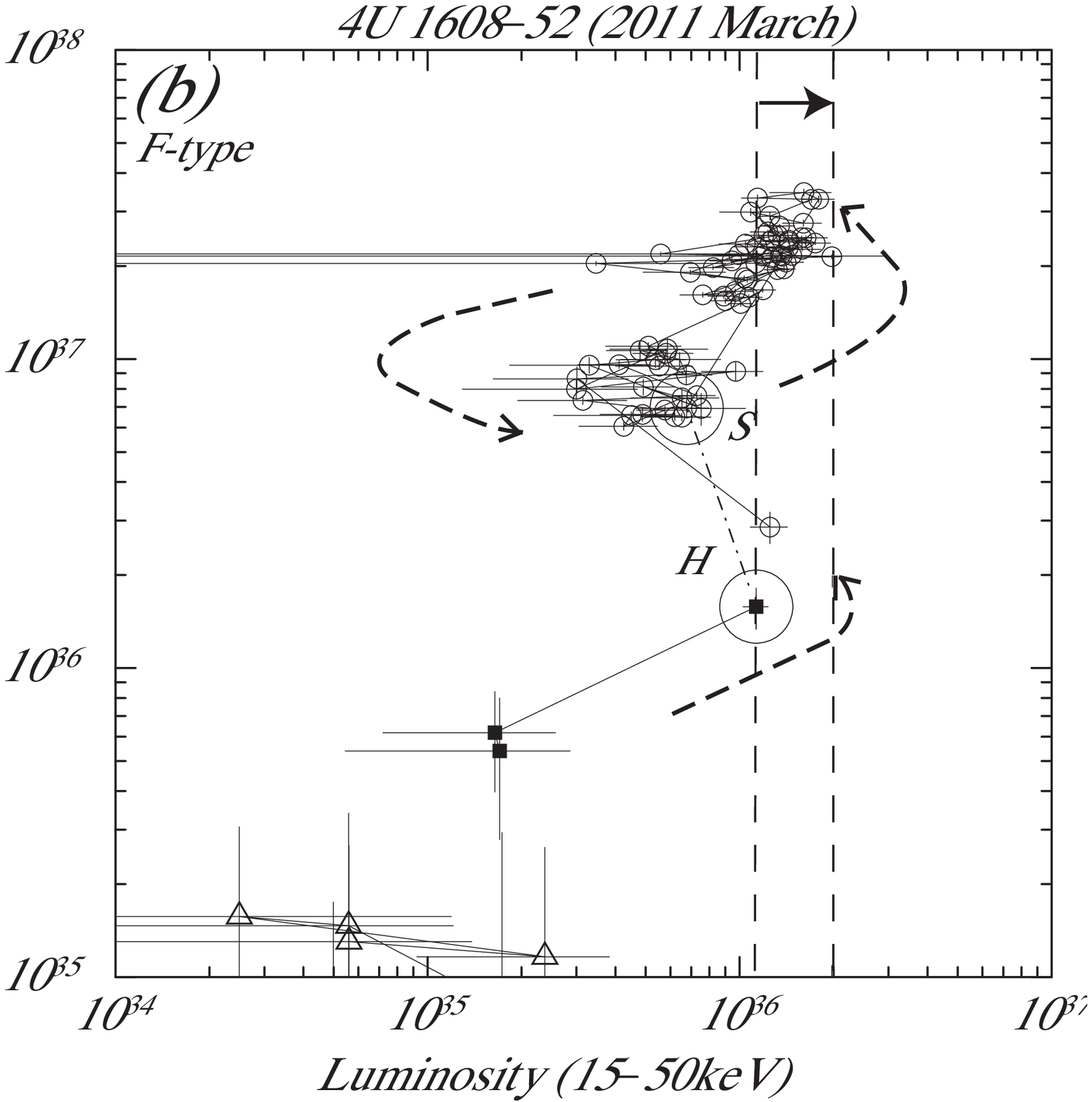}
    \FigureFile(70mm,){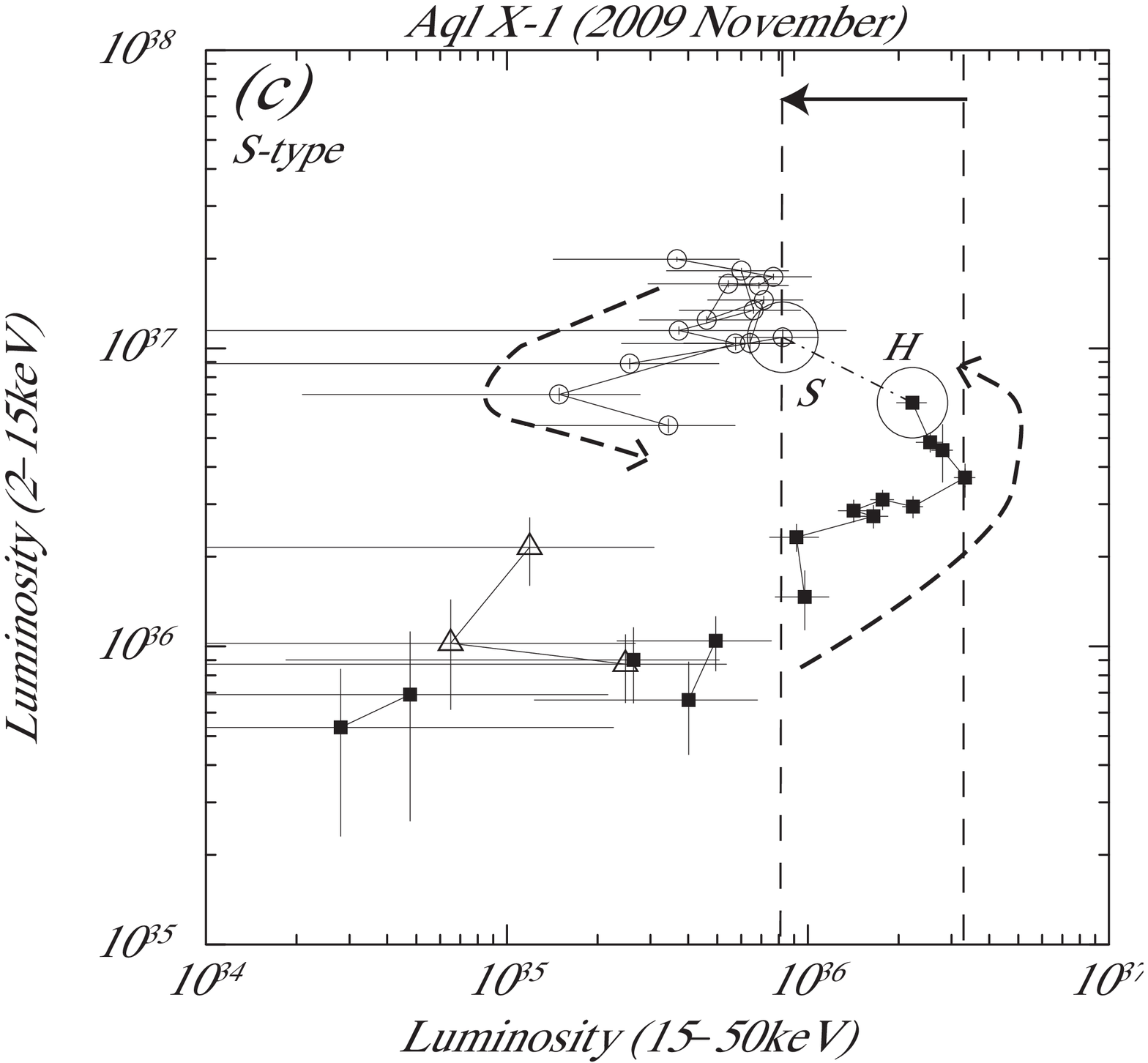}
    \FigureFile(64mm,){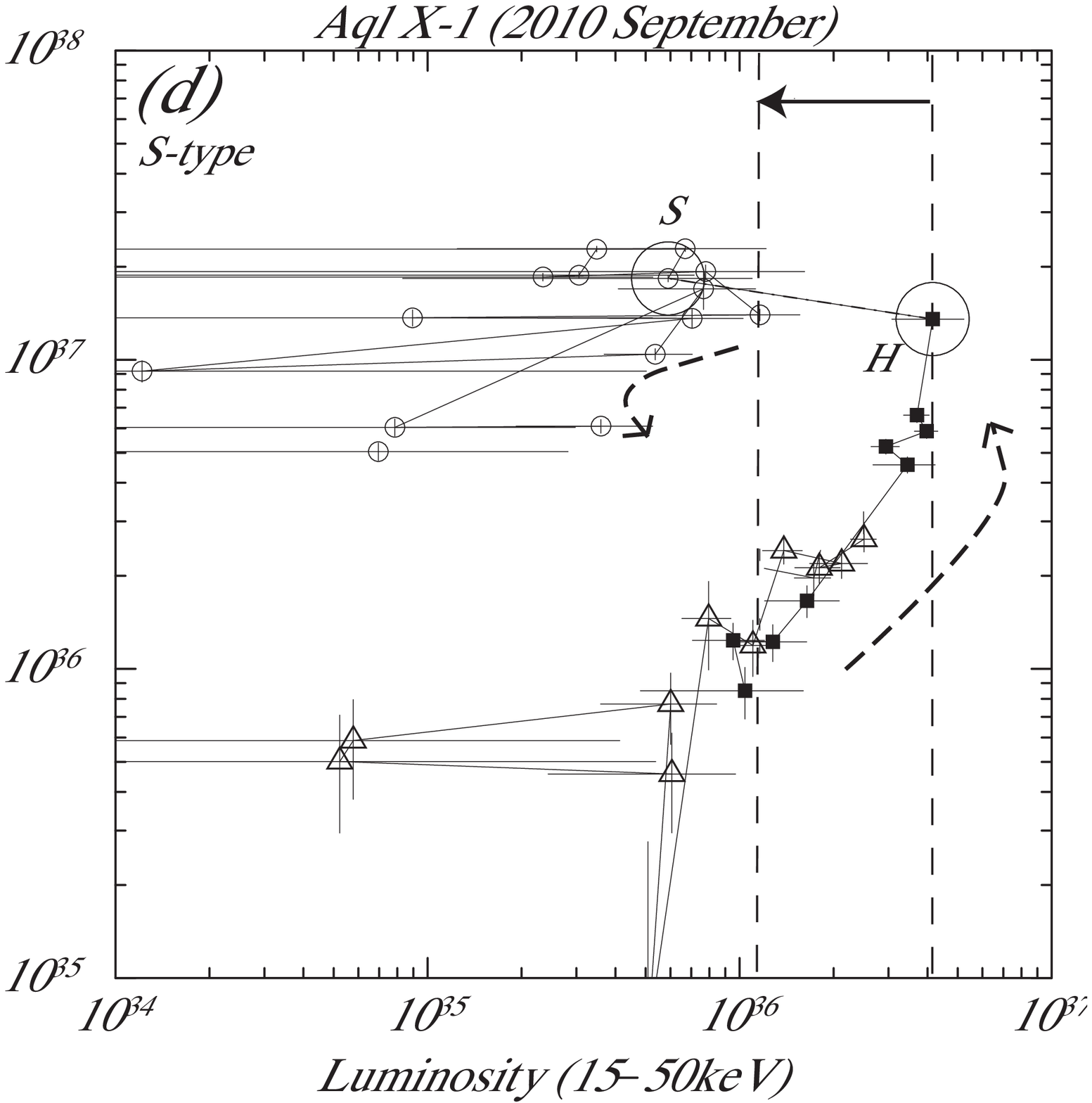}
    \FigureFile(70mm,){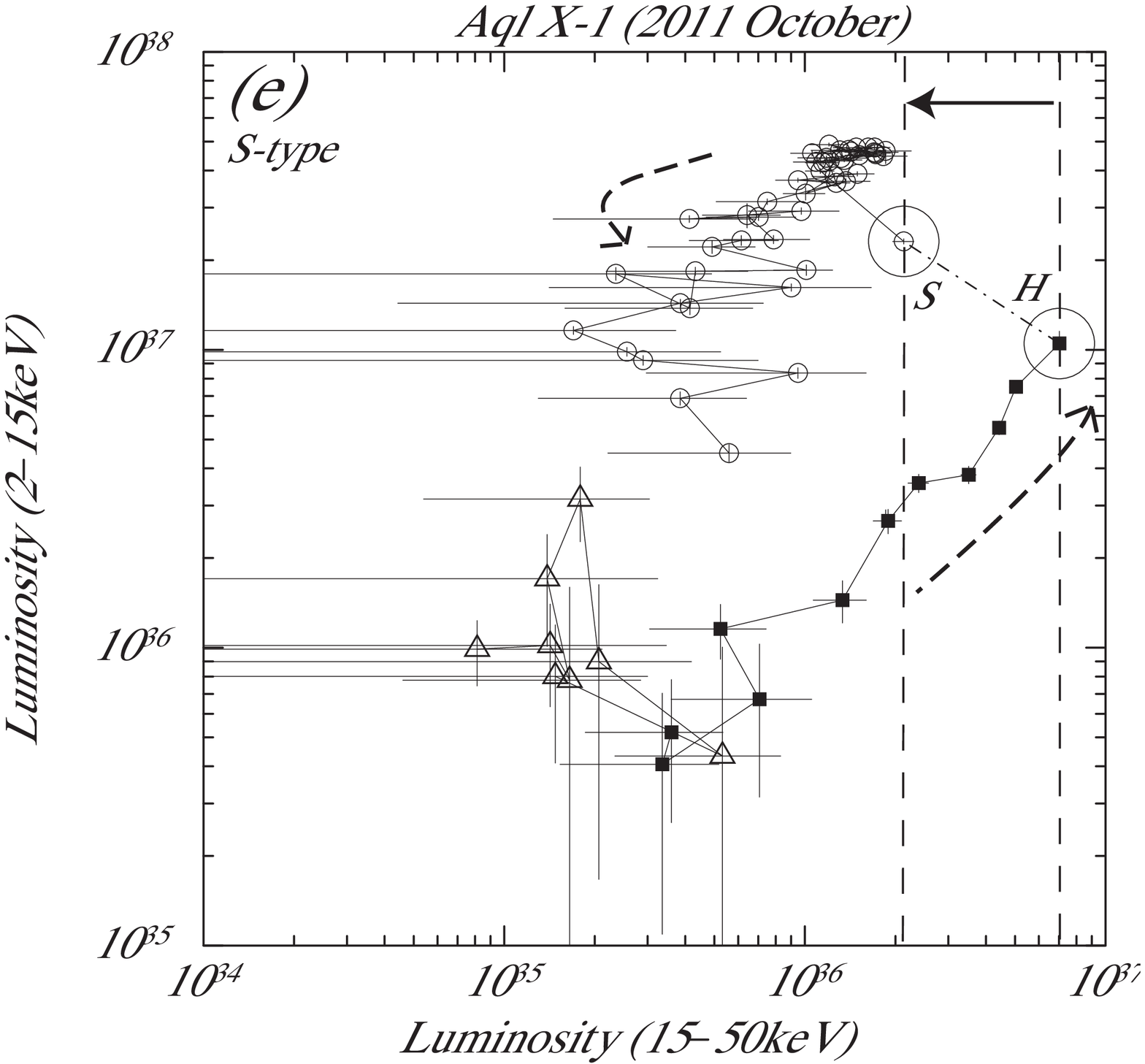}
  \end{center}
  \caption{GSC (2--15~keV) and BAT (15--50~keV) intensity--intensity diagram.
\textcolor{black}{
Error bars represent 1-$\sigma$ statistical uncertainties.
}
Open triangles, filled squares, open circles represent data
before the outburst start, during the initial hard state,
and during the soft state, respectively.
Furthermore, we connected the data points in the same state
along the time evolution.
Labels of ``H'', and ``S'' represent the data at just
before/after the hard-to-soft transition.
\textcolor{black}{
We connected data points before and after the hard-to-soft transition (H and S)
with the dash-dot line.
}
The dashed lines with arrows indicate the time sequence of each outburst.
The two vertical dashed lines indicate the 15--50~keV peak luminosities
during each hard/soft state.
The top horizontal arrow indicates the direction of the change from
the hard state to the soft state.}
\label{fig3}
\end{figure*}

Figure~\ref{fig2} shows GSC light curves, BAT light curves,
and the hardness ratios (BAT/GSC) for the five outbursts.
First, we estimated the outburst onset from the GSC light curves
with a 1-day time bin.
The archived GSC light-curve data include a systematic error
of $\sim$ 5mCrab for the imperfection in the background subtraction.
We confirmed that the systematic error does not affect the 
results of the following analysis significantly.
Once outbursts started, the flux increased steadily from nominal
low level below 10~mCrab towards the peak over 100~mCrab.
When the trend of source luminosity before outburst onset was constant
or decreasing, the outburst onset can be determined clearly.
When the source luminosity before outburst onset was increasing,
we determine the onset in the following way.
The outburst typically shows a sharp exponential rise, whose slope is
different from the persistent component.
However, the rise curve sometimes has more than one slope even in  the
rising period, then
we cannot determine the onset just by the changing point of the slopes.
Therefore in this paper, 
we define the outburst onset as the start-point
of the increase trend towards the peak.
In order to determine the start-point, we fitted light curves around
the outburst onset with two exponential functions, and define
the ``cross point'' of the two exponential functions as the
start-point of the outburst.
The results are shown in the Appendix.

We next estimate the time and the luminosity when the hard-to-soft
transition occurs from data bins of the ``last hard state'' and the
``first soft state''.
We employed the center of two data bins as
``transition time (time when transition occurs)'' 
and their time interval as the error on it.
We also defined their average luminosity as the
``transition luminosity''.
Thus, the ``pre-transition time'' is defined by the duration from
the ``outburst onset'' to the ``transition time''.
The derived epoch of the ``outburst onset'',
``pre-transition time'', and ``transition luminosity''
for each outburst are summarized in table~1.

\begin{table*}
\caption{Parameters of five outbursts observed by MAXI/GSC and Swift/BAT.}
\label{tab1}
\begin{center}
\begin{tabular}{lccccc}
\hline
& \multicolumn{2}{c}{4U~1608$-$52} & \multicolumn{3}{c}{Aql~X-1} \\
& (2010 Mar) & (2011 Mar) & (2009 Nov) & (2010 Sep) & (2011 Oct) \\
\hline
Outburst onset
\footnotemark[$*$](MJD)
&
\textcolor{black}{$55253.5\pm0.5$} & \textcolor{black}{$55641.5^{+0.8}_{-0.6}$} &
\textcolor{black}{$55134.4\pm0.6$} & \textcolor{black}{$55440.7\pm1.0$} &
\textcolor{black}{$55842.2^{+1.0}_{-0.7}$}\\
\hline
Transition time\footnotemark[$\dagger$](MJD) &
$55255.5\pm2.0$
& $55644.0\pm0.5$ & $55153.0\pm0.5$ & $55451.0\pm0.5$ & $55856.0\pm0.5$ \\
\, \, Last hard state & & & & & \\
\, \, \, \, time\footnotemark[$\ddagger$](MJD)
& \textcolor{black}{$55253.5\pm0.5$} & $55643.5\pm0.5$ & $55152.5\pm0.5$ & $55450.5\pm0.5$ & $55855.5\pm0.5$ \\
\, \, \, \, Hardness ratio\footnotemark[$\S$]
& \textcolor{black}{$2\pm1$} & $1.3\pm0.2$ & $0.66\pm0.08$ & $0.6\pm0.1$ & $1.30\pm0.07$ \\
\, \, \, \, Luminosity\footnotemark[$\l$]
& \textcolor{black}{$0.03\pm0.01$} & $0.16\pm0.02$ & $0.66\pm0.03$ & $1.35\pm0.04$ & $1.05\pm0.04$ \\  
\, \, First soft state & & & & & \\
\, \, \, \, time\footnotemark[$\ddagger$](MJD)
& $55257.5\pm0.5$ & $55644.5\pm0.5$ & $55153.5\pm0.5$ & $55451.5\pm0.5$ & $55856.5\pm0.5$ \\
\, \, \, \, Hardness ratio\footnotemark[$\S$]
& \textcolor{black}{$<0.4$} & $0.19\pm0.04$ & $0.15\pm0.05$ & $0.06\pm0.05$ & $0.18\pm0.01$ \\
\, \, \, \, Luminosity\footnotemark[$\l$]
& $0.09\pm0.01$ & $0.69\pm0.05$ & $1.09\pm0.07$ & $1.83\pm0.05$ & $2.31\pm0.05$ \\
\hline
Pre-transition time\footnotemark[$\#$](days)&
\textcolor{black}{$3.5\pm1.0$} &  \textcolor{black}{$2.5^{+1.1}_{-1.3}$} &
\textcolor{black}{$18.6\pm1.1$} &
\textcolor{black}{$10.3\pm1.5$} & \textcolor{black}{$13.8^{+1.2}_{-1.5}$}\\
Transition luminosity\footnotemark[$**$] &
\textcolor{black}{$0.06\pm0.03$}  & $0.4\pm0.3$ & $0.9\pm0.2$ &
$1.6\pm0.2$ & $1.7\pm0.6$\\
Soft/Hard state at 15--50~keV max.\footnotemark[$\dagger\dagger$]
& soft state & soft state & hard state &  hard state & hard state \\
Type & F & F & S & S & S \\
\hline
\\
\multicolumn{6}{@{}l@{}}{\hbox to 0pt{\parbox{180mm}{\footnotesize
\par\noindent
\footnotemark[$*$]
\textcolor{black}{
See the Appendix.
}
\par\noindent
\footnotemark[$\dagger$]
\textcolor{black}{
The ``transition time'' is defined by the middle of the
``last hard state'' and the ``first soft state'', 
and the error is defined by the span between
the both data points.
}
\par\noindent
\footnotemark[$\ddagger$]
\textcolor{black}{
All errors of the ``time [MJD]'' are $\pm0.5$~d,
indicating that the data points are for 1-day average.
}
\par\noindent
\footnotemark[$\S$]
Hardness ratio means 15--50~keV/2--15~keV.
\par\noindent
\footnotemark[$\l$]
\textcolor{black}{
Luminosity in 2--15~keV band in the unit of $10^{37}$~erg~s$^{-1}$.
The errors in "Luminosity" are 1-$\sigma$ errors.
}
\par\noindent
\footnotemark[$\#$]
``Pre-transition time'' means the initial hard-state duration from
 ``Outburst onset'' to the ``Transition time''.
\par\noindent
\footnotemark[$**$]
``Transition luminosity'' is an average luminosity from ``Last hard state''
to ``First soft state'' in 2--15~keV band in unit of $10^{37}$~erg~s$^{-1}$.
\par\noindent
\footnotemark[$\dagger\dagger$]
Whether soft state or hard state when the 15--50~keV luminosity reached 
the maximum.
}\hss}}
\end{tabular}
\end{center}
\end{table*}

To inspect the relative intensity variations between the GSC and the BAT bands,
GSC--BAT intensity--intensity diagrams from 10~d
before the outburst
onset to the end of the soft state are plotted in figure~3, 
where fluxes in the GSC and BAT light curves are converted to the luminosities
in the 2--15 keV and in the 15--50 keV, respectively, assuming that
the spectrum is Crab-like (Kirsch et al. 2005).

These intensity--intensity diagrams are obviously classified into two
types.
One type behaves as the two intensities in the different
energy bands of 2--15~keV and the 15--50~keV become the maximum almost
simultaneously in the soft state after the hard-to-soft transition
occurred (figures~\ref{fig3}a, \ref{fig3}b).
\textcolor{black}{
The other type behaves as the 15--50~keV intensity reaches the maximum
earlier than the 2--15~keV intensity
(figures~\ref{fig3}c, \ref{fig3}d, \ref{fig3}e).
Then, the 15--50~keV intensity during the soft state does not exceed
the maximum just before the hard-to-soft transition state,
whereas the 2--15~keV intensity
reaches its maximum after the hard-to-soft transition. 
}
These two types are apparently related with the
pre-transition time as shown in table~\ref{tab1}.
\textcolor{black}{
The pre-transition time is shorter than 5~d in the former type
but longer than 9~d in the latter type for the five outbursts
observed by GSC.
}
We therefore name the former as Fast(F)-type and the latter as Slow(S)-type.

\subsection{RXTE-ASM, Swift-BAT Light Curves and the Hardness Evolution}

\begin{figure*}
  \begin{center}
    \FigureFile(80mm,){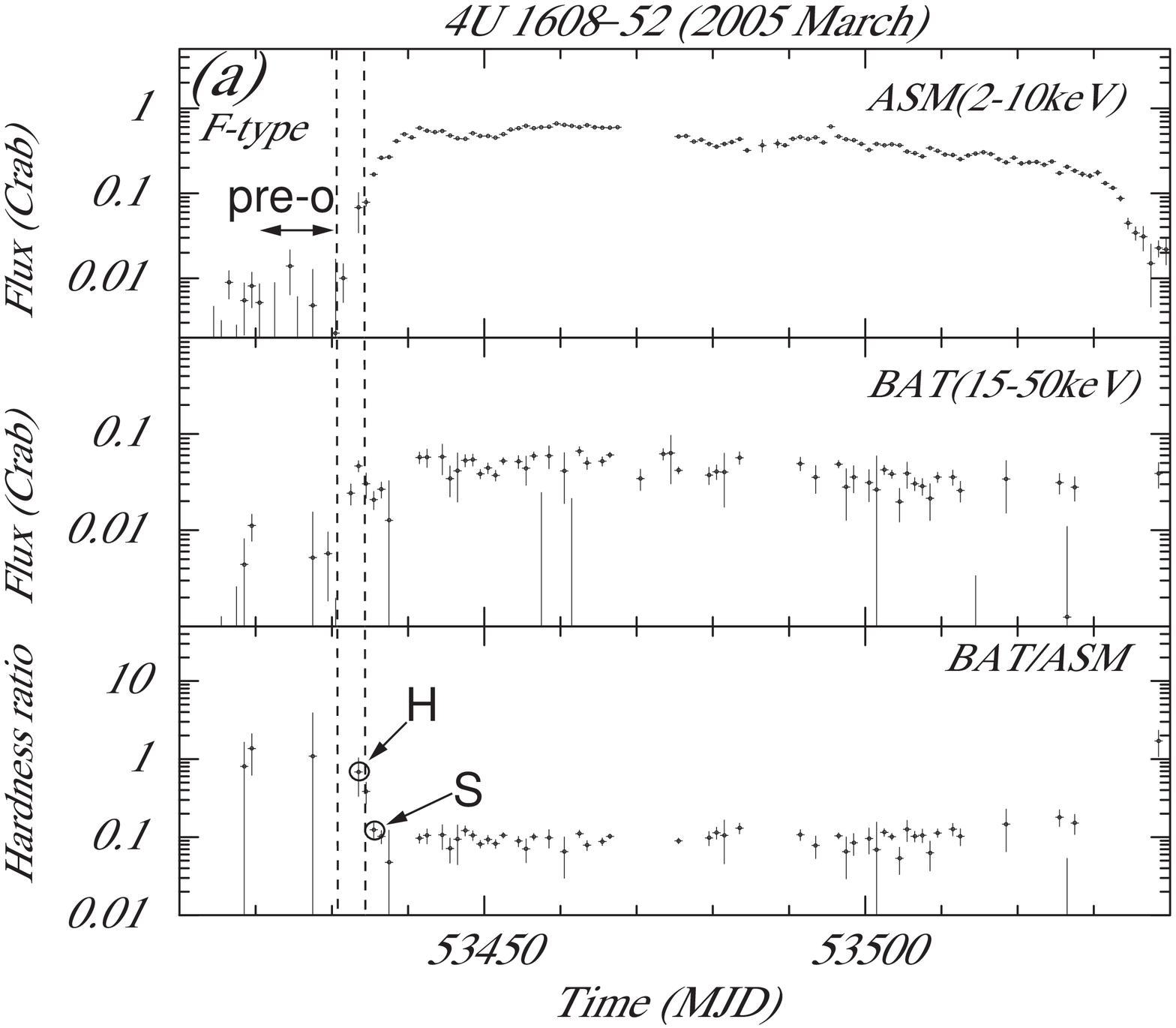}
    \FigureFile(80mm,){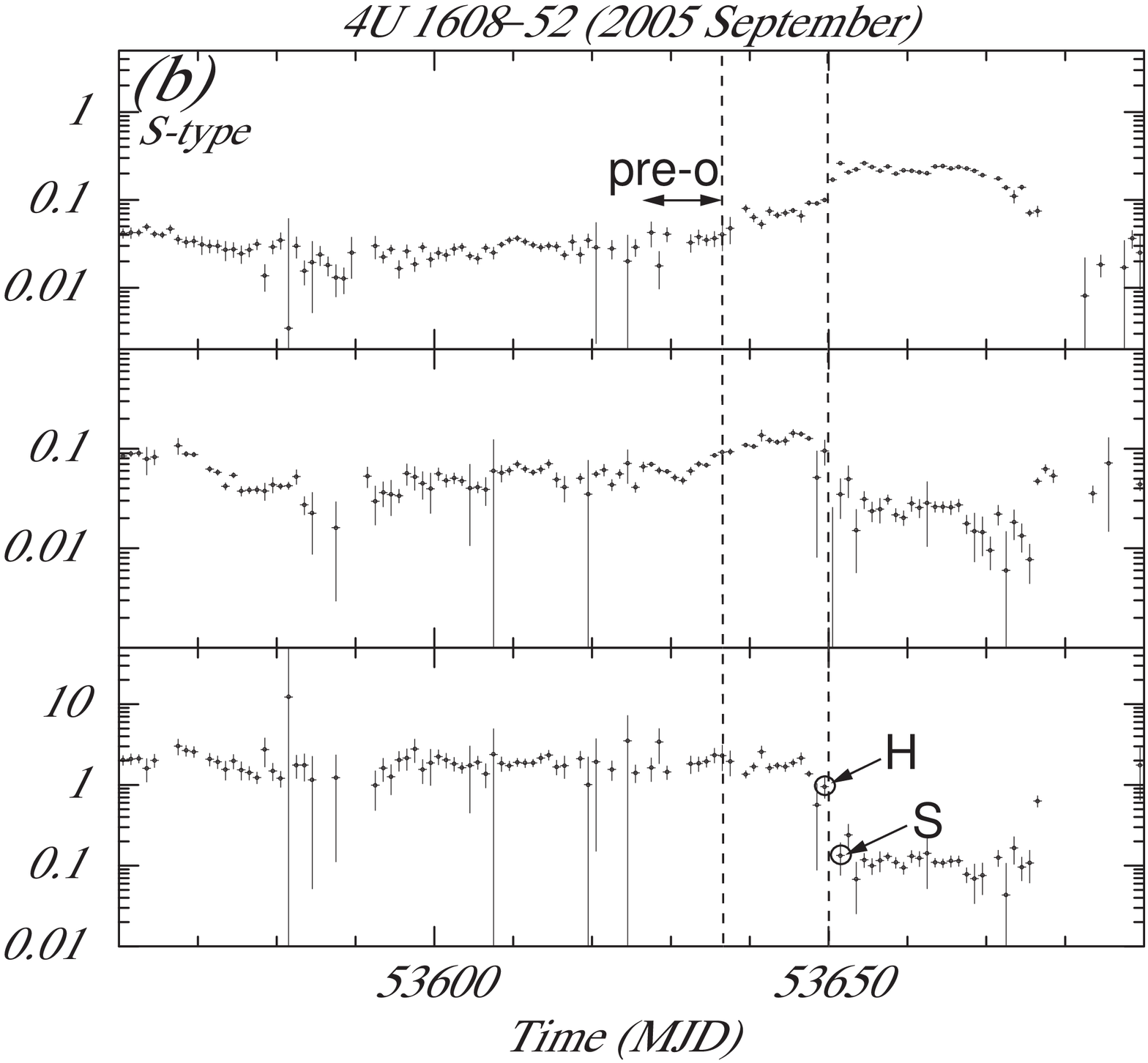}
    \FigureFile(80mm,){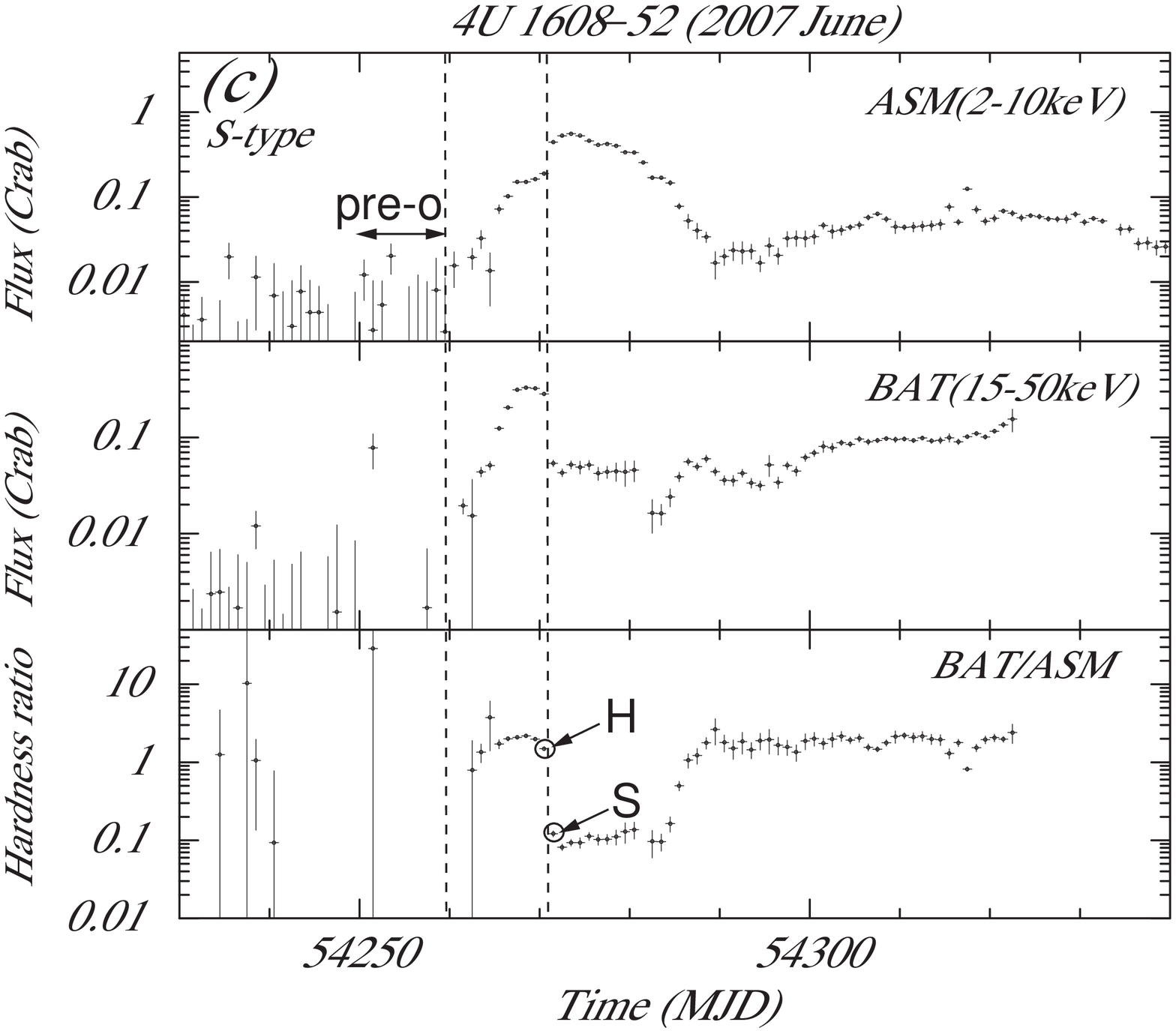}
    \FigureFile(80mm,){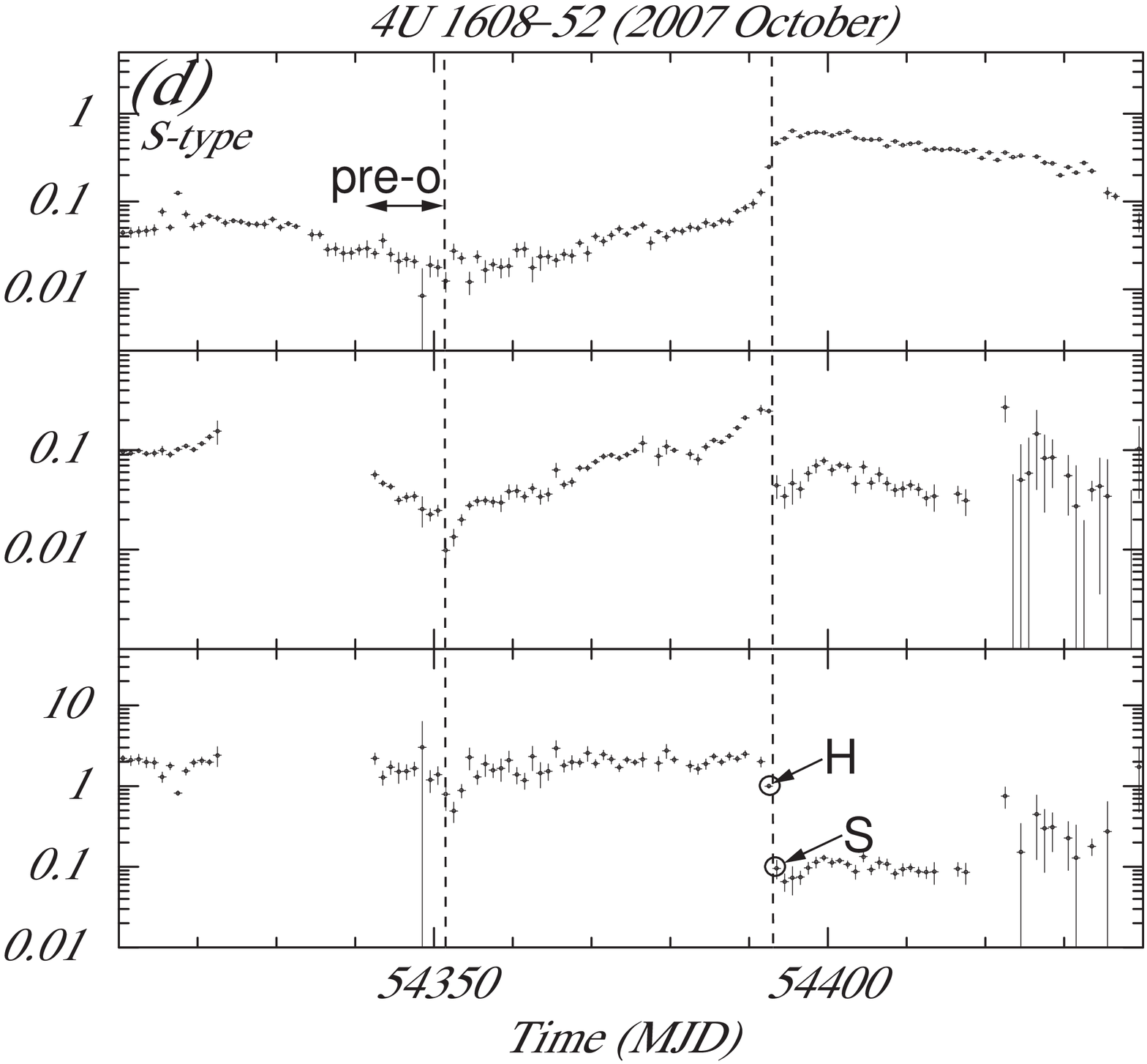}
    \FigureFile(80mm,){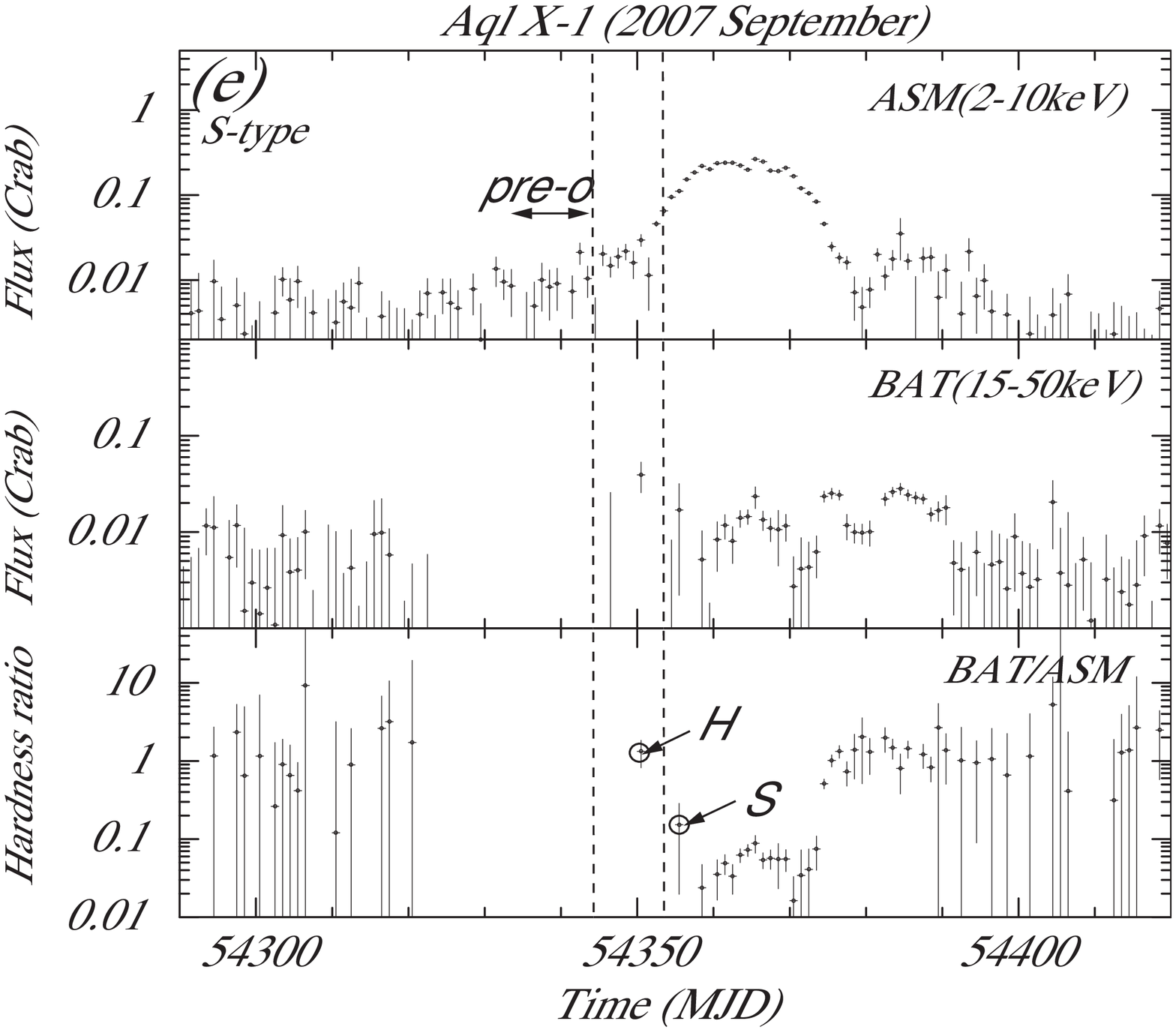}
  \end{center}
  \caption{RXTE-ASM light curves (2--10~keV),
Swift-BAT light curves (15--50~keV),
and the hardness ratios (BAT/ASM) of five outbursts.
The labels and 
\textcolor{black}{errors}
are the same as in figure~\ref{fig2}.
}
\label{fig4}
\end{figure*}

\begin{figure*}
  \begin{center}
    \FigureFile(70mm,){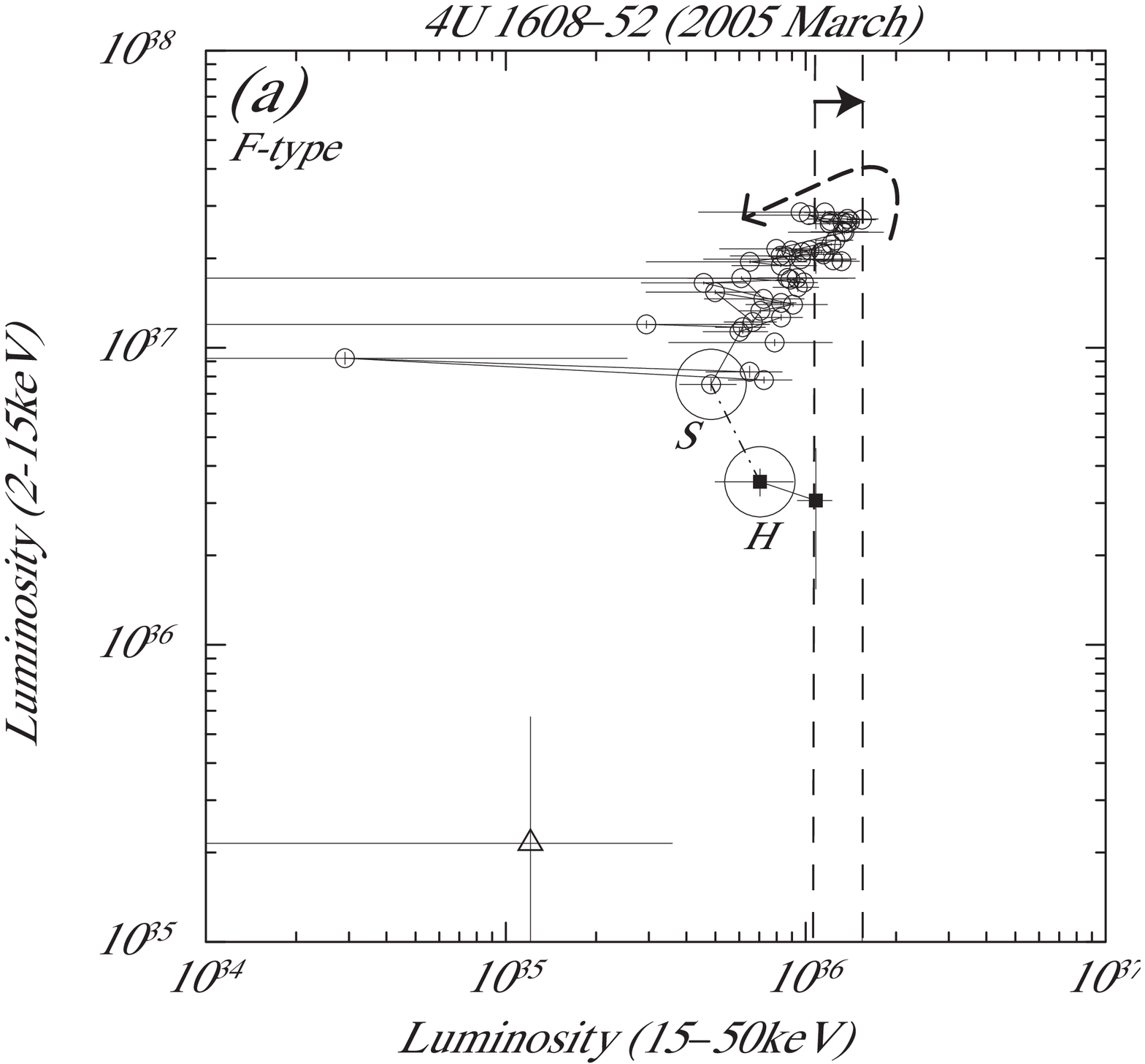}
    \FigureFile(64mm,){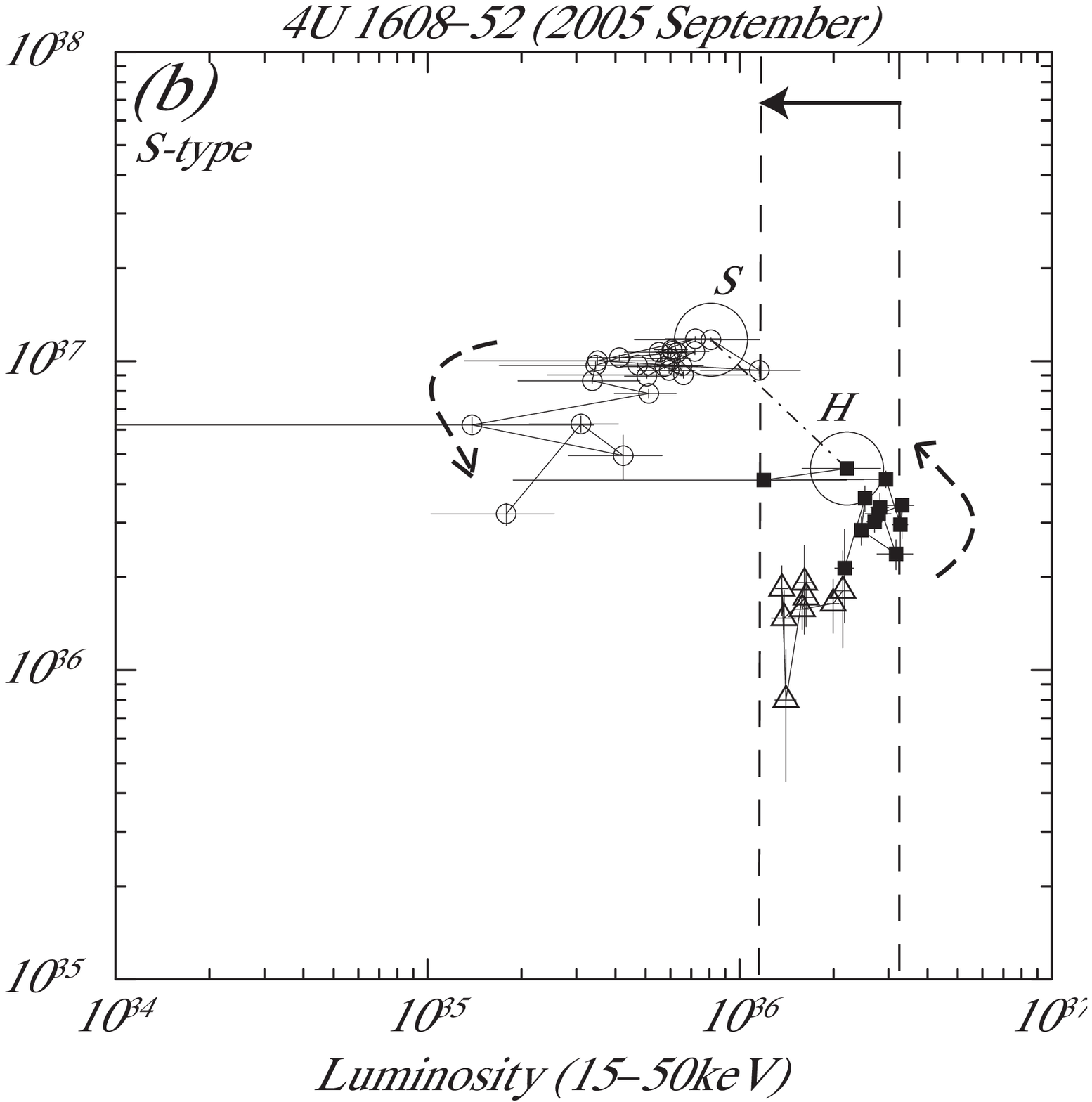}
    \FigureFile(70mm,){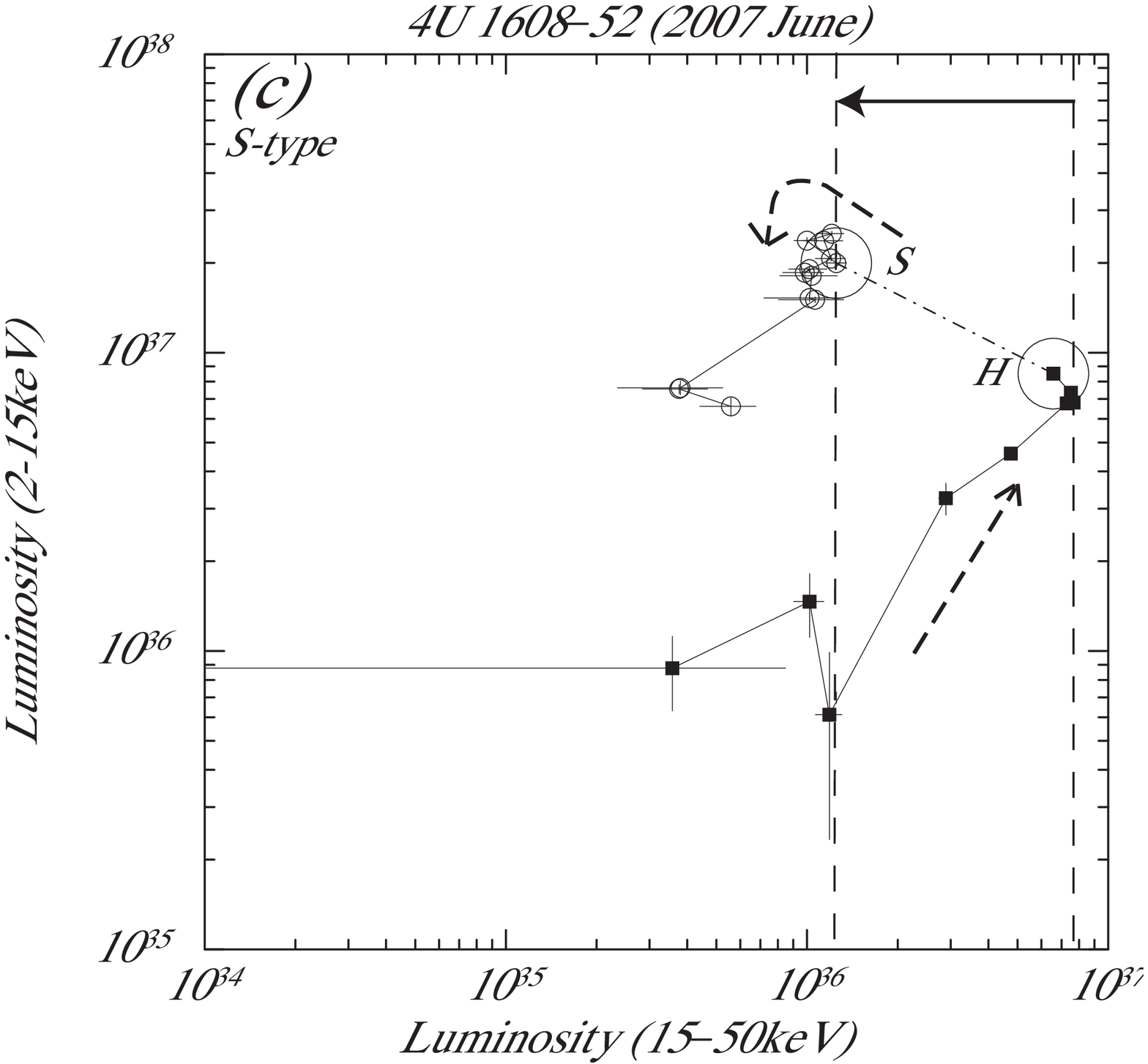}
    \FigureFile(65mm,){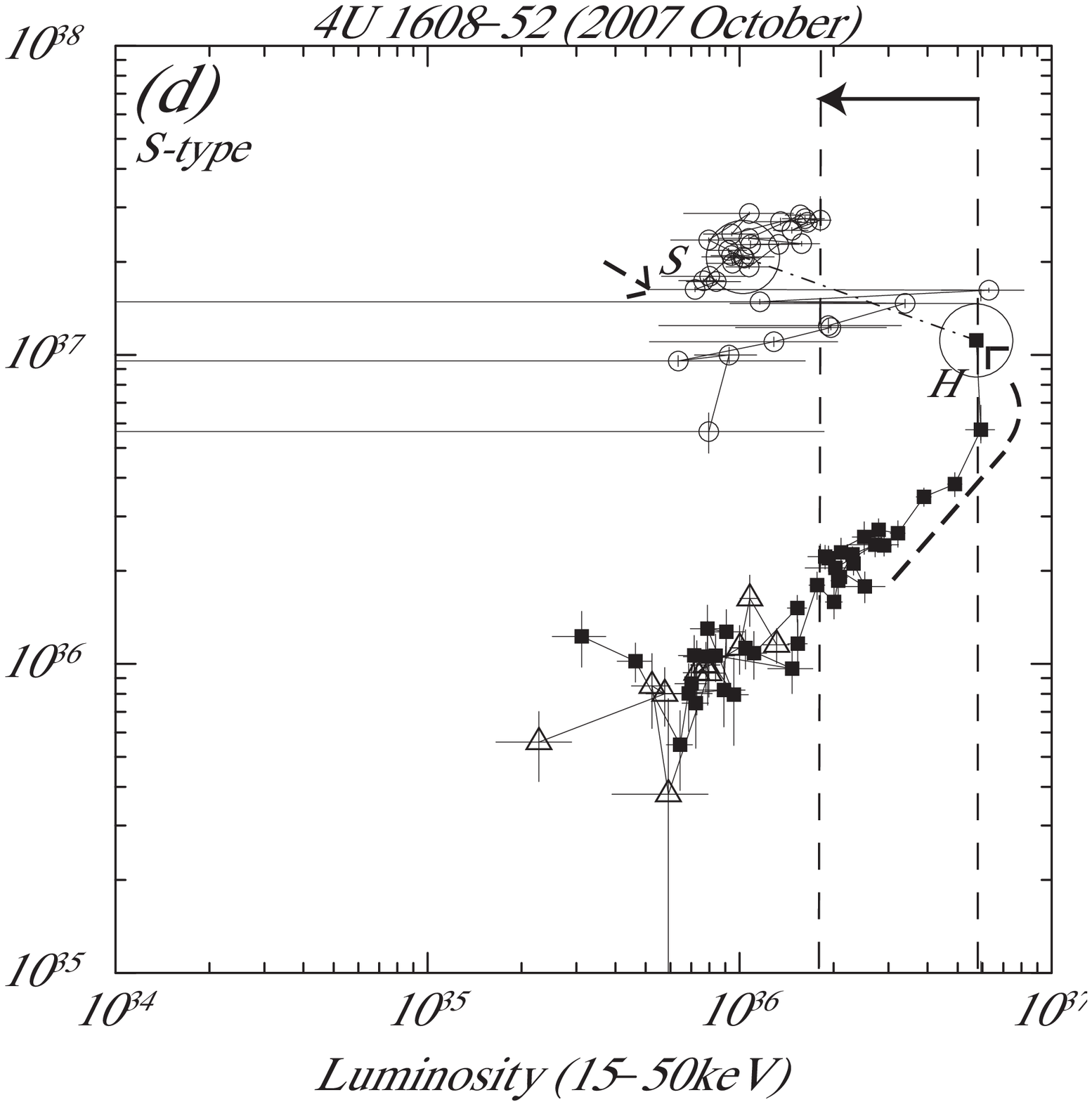}
    \FigureFile(70mm,){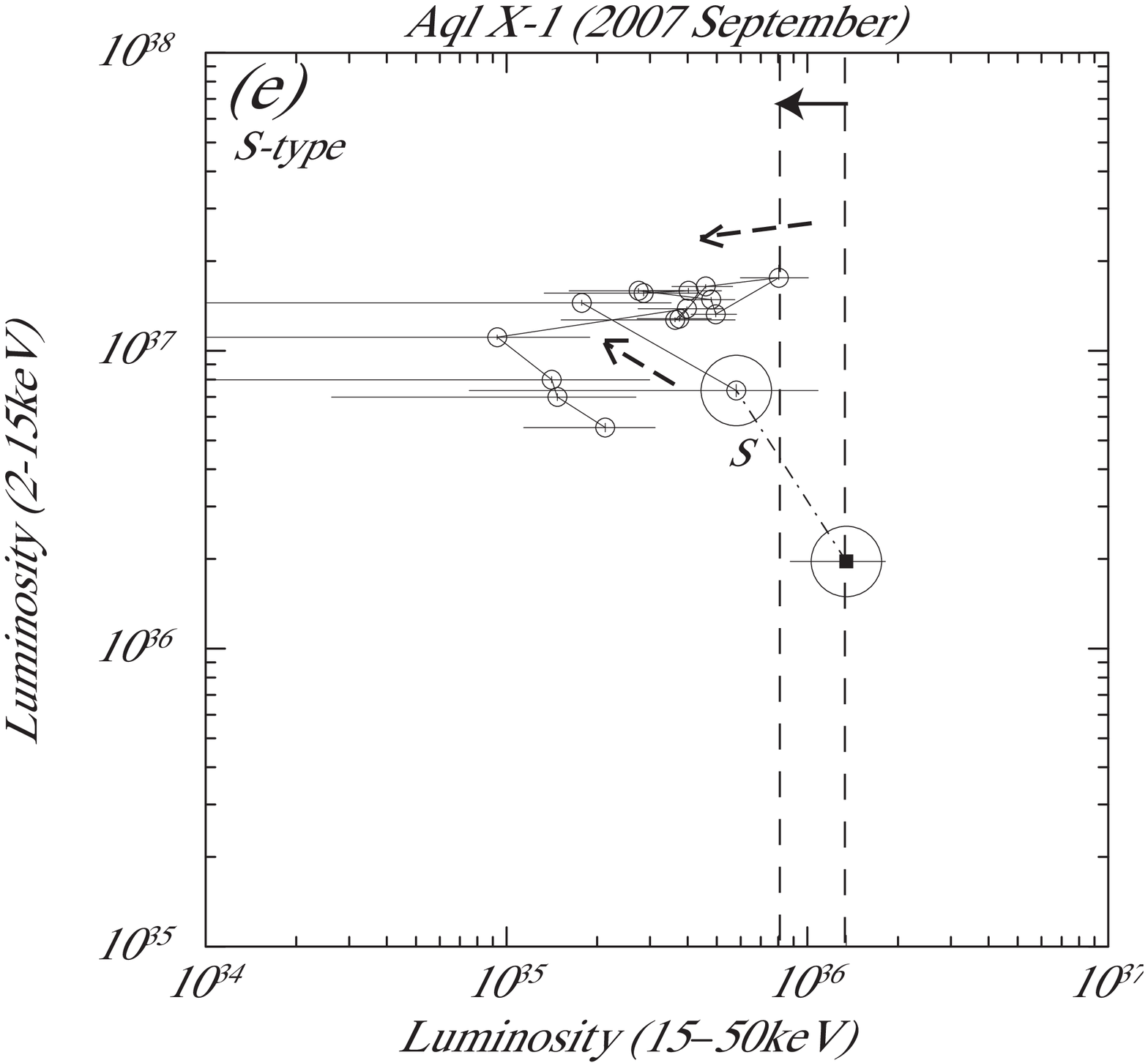}
  \end{center}
  \caption{Same as figure~\ref{fig3}, but using ASM and BAT data.}
\label{fig5}
\end{figure*}

To confirm the results obtained from the MAXI-GSC and Swift-BAT light curves
in the previous subsection, we also analyzed outbursts from the same sources
observed with the ASM (All Sky Monitor: Levine et al. 1996) inboard 
RXTE (Rossi X-ray Timing Explore: Bradt et al. 1993) 
in 2--10~keV.
The ASM data are obtained from the archived results provided 
by the RXTE-ASM teams at MIT and NASA/GSFC. 
\footnote{$<$http://xte.mit.edu/$>$.}
We found four outbursts from 4U~1608$-$52 and one outburst from Aql~X-1
since the Swift started the operation in 2004 November until the MAXI
started in 2009 August. 
We performed the same analysis procedure as applied for the GSC and BAT data
in the previous subsection.
Figure~4 shows the ASM light curves in 2--10~keV, the BAT light curves
in 15--50~keV, and the hardness ratios of BAT/ASM for the five outbursts,
where ASM data are converted to a flux in Crab unit using the nominal
relation of 1~Crab = 75~counts s$^{-1}$ for ASM.
\footnote{$<$http://xte.mit.edu/XTE/ASM\_lc.html$>$.}

 We also examined the hardness-ratio distribution as has been shown
in figure~1. 
\textcolor{black}{
We used all of the data points obtained from 2005 February 13 (MJD=53414)
to 2011 July 31 (MJD=55773) after Swift started the operation.
} 
We fitted the histogram with
a model of two log-normal distribution functions
and determined
the soft-hard threshold at the center of the two peaks obtained
from the best-fit parameters.
The center values of the Gaussian ($C_1$ and $C_2$) are
$\log_{10} C_1 = 0.23 \pm 0.01, \log_{10} C_2 = -0.96 \pm 0.02$ and
$\log_{10} C_1 = 0.13 \pm 0.02, \log_{10} C_2 = -1.1 \pm 0.2$,
for 4U~1608$-$52 and Aql~X-1, respectively.
The obtained thresholds $10^{(\log_{10} C_1+ \log_{10} C_2)/2}$, 
are 0.43 and 0.34, respectively.
Figure~5 shows the ASM--BAT intensity--intensity diagrams
representing the relations between the 2--15~keV and the 15--50~keV
intensities from 10~d before the outburst onset to the end of the soft state.
In the outburst of 4U~1608$-$52 in 2007 October (figure~4),
a large flux increase in the 15--50~keV band was observed in the soft state,
which seems to be the second peak in the BAT band.

\begin{table*}
\caption{Parameters of five outbursts observed
by RXTE/ASM and Swift/BAT.\footnotemark[$*$]}
\label{tab2}
\begin{center}
\begin{tabular}{lccccc}
\hline
& \multicolumn{4}{c}{4U~1608$-$52} & Aql~X-1 \\
& (2005 Mar) & (2005 Sep) & (2007 Jun) & (2007 Oct) & (2007 Sep) \\
\hline
\textcolor{black}{
Outburst onset (MJD)} &
\textcolor{black}{$53430.7^{+1.2}_{-0.7}$} & \textcolor{black}{$53637\pm2$} & 
\textcolor{black}{$54259\pm1$} & \textcolor{black}{$54352^{+8}_{-2}$} &
\textcolor{black}{$54345\pm3$} \\
\hline
Transition time (MJD) &
\textcolor{black}{$53434.5\pm1.0$} & $53650.5\pm1.0$ & $54271.0\pm0.5$ & $54393.0\pm0.5$ &
$54353.5\pm2.5$ \\
\, \, Last hard state & & & & & \\
\, \, \, \, time (MJD)
& \textcolor{black}{$53433.5\pm0.5$} & $53649.5\pm0.5$ &
$54270.5\pm0.5$ & $54392.5\pm0.5$ & $54350.5\pm0.5$ \\
\, \, \, \, Hardness ratio
& \textcolor{black}{$0.7\pm0.4$} & $1.0\pm0.3$ & $1.50\pm0.09$ &
$1.00\pm0.05$ & $1.3\pm0.5$ \\
\, \, \, \, Luminosity
& \textcolor{black}{$0.3\pm0.2$} & $0.45\pm0.02$ &
$0.85\pm0.04$ & $1.11\pm0.03$ & $0.20\pm0.03$ \\  
\, \, First soft state & & & & & \\
\, \, \, \, time (MJD)
& $53435.5\pm0.5$ & $53651.5\pm0.5$ &
$54271.5\pm0.5$ & $54393.5\pm0.5$ & $54355.5\pm0.5$ \\
\, \, \, \, Hardness ratio
& $0.12\pm0.03$ & $0.13\pm0.06$ & $0.12\pm0.01$ &
$0.10\pm0.03$ & $0.2\pm0.1$ \\
\, \, \, \, Luminosity 
& $0.75\pm0.04$ & $1.17\pm0.02$ &
$2.00\pm0.05$ & $2.07\pm0.05$ & $0.74\pm0.04$ \\
\hline
Pre-transition time (days) &
\textcolor{black}{$3.8^{+1.7}_{-2.2}$} &  \textcolor{black}{$13.5\pm3$} &
\textcolor{black}{$12\pm1.5$} & \textcolor{black}{$41^{+2.5}_{-8.5}$}  &
\textcolor{black}{$8.5\pm5.5$} \\
Transition luminosity &
$0.6\pm0.2$  & $0.8\pm0.4$ & $1.4\pm0.6$ & $1.6\pm0.5$ & $0.5\pm0.3$\\
Soft/Hard state at 15--50~keV max.
& soft state & hard state &  hard state & hard state
& hard state \\
Type & F & S & S & S & \textcolor{black}{S} \\
\hline
\\
\multicolumn{6}{@{}l@{}}{\hbox to 0pt{\parbox{180mm}{\footnotesize
\par\noindent
\footnotemark[$*$] The same table as table~1 but for ASM and BAT data.
}\hss}}
\end{tabular}
\end{center}
\end{table*}

Table~2 summarizes the parameters of the five outbursts
by ASM and BAT obtained in the same manner as in table~1.

\textcolor{black}{
Since the Aql~X-1 outburst in 2007 September has a large data gap
in the BAT data around the ``transition time (MJD)'',
``pre-transition time (days)'' has large errors.
}
All the obtained results confirmed that the outbursts detected
by ASM also show the same features of either S-type or F-type as
seen in the five outbursts detected by GSC as described in subsection 2.1.
\textcolor{black}{
The pre-transition times in the F-types are shorter than that in the S-type.
}
\subsection{Relation of Pre-outburst Luminosity, Transition Luminosity on Transition Behavior}
\label{Luminosity before Outburst section}

\begin{figure}
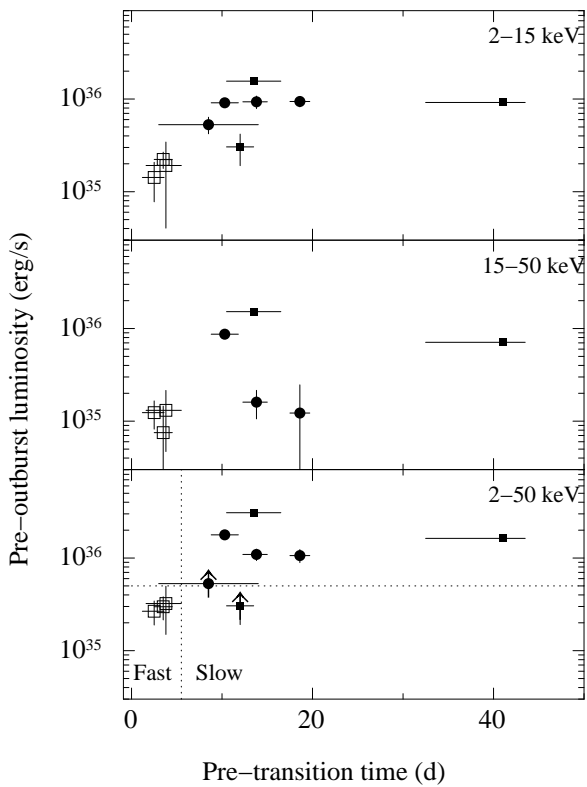

  \begin{center}
    \FigureFile(80mm,){fig6.eps}
  \end{center}
  \caption{Relation between pre-outburst luminosity in 2--15~keV (top),
15--50~keV (middle), 2--50~keV (bottom) 
and ``pre-transition time''.
The errors in ``pre-transition time'' 
are obtained in the same manner as in table~1.
The errors in ``pre-outburst luminosity'' are 1-$\sigma$ errors
\textcolor{black}{
(cf. footnote 5).
}
Squares are for 4U~1608$-$52 and circles for Aql~X-1.
Filled and open marks represent S-type and F-type, respectively.}
\label{fig6}
\end{figure}

To find the root causes for the two behaviors of S-type and
F-type, we here investigate parameters that have a correlation with
the pre-transition time.

Firstly, we examined the source luminosities before the outburst
onset.
The source luminosities before the outburst onsets changed more than 
one order of magnitude in several time scales (5--100~d)
even in the quiescent phase.
In calculating ``pre-outburst luminosity'', 
 \footnote{
\textcolor{black}{
We adopted all the public data of MAXI/GSC and RXTE/ASM through the
present analysis. However,} 
\textcolor{black}{
the weaker intensities probably include some
background component. Thus ``pre-outburst luminosities'' for the
F-types may be contaminated with the background component considerably
and may be upper limit. Nevertheless, the present conclusion is not affected.
}
}
we adopted 10~d interval,
\textcolor{black}{
which according to average
transition time scale.}
Figure~\ref{fig6}
shows the correlations of the 2--15~keV luminosity by GSC
or ASM, the 15--50~keV luminosity by BAT, and their sum of the
2--50~keV band to the pre-transition time.
\textcolor{black}{
For the five outbursts observed with both the GSC and the ASM
after the MAXI started the operation, we calculated pre-outburst
luminosities in 2-15~keV band for both instruments and confirmed that
the results are consistent with each other.
Here, we employ GSC data because the statistical accuracy of
the GSC data is better than that of the ASM.
}
The pre-outburst luminosity of S-type tends to be higher
than that of F-type.

In the middle panel of figure~6, two data points
\textcolor{black}{
(4U~1608$-$52 outbursts in 2007 June and Aql~X-1 outburst
in 2007 September)
}
are not plotted
because the public BAT light curves during these two outbursts
have data gaps or data points with negative count rates. 
Therefore, the average luminosities were not able to be obtained.
Although these data are omitted in the middle panel
we showed them in the bottom panel
using the lower limits.

Finally, We examined the correlation between the transition luminosity
(tables~\ref{tab1} and \ref{tab2}) and the pre-transition time.
Figure~\ref{fig7} shows the obtained relation.
The F-type points are in the left-down section while S-type points
locate in the right-upper section.
The transition luminosity of S-type is larger than that of F-type.

\begin{figure}
  \begin{center}
    \FigureFile(80mm,){fig7.eps}
  \end{center}
  \caption{Relation between transition luminosity and pre-transition time.
\textcolor{black}{
The errors are obtained in the same manner as in table~1.}
Symbols are the same as those of figure~6.}
\label{fig7}
\end{figure}

\section{Discussion}

\begin{figure*}
  \begin{center}
    \FigureFile(110mm,){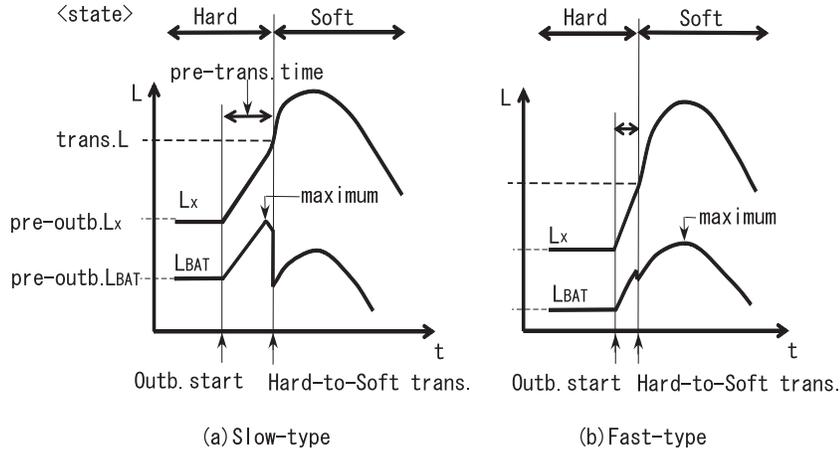}
  \end{center}
  \caption{
Schematic drawings of 2--15~keV and 15--50~keV light curves
    (L$_{\rm X}$ and L$_{\rm BAT}$) in each of S-type and F-type
    outbursts.  Four differences in pre-transition time, transition
    luminosity, pre-outburst luminosity, and location of the
    15--50~keV maximum luminosity between the S-type and the F-type
    are illustrated.}
\label{fig8}
\end{figure*}

We analyzed the initial rising behaviors of ten outbursts from Aql~X-1
and 4U~1608$-$52 using data taken by MAXI/GSC, Swift/BAT, and
RXTE/ASM.
We have found the two types of hard-to-soft state transitions.
One type behaves as the 15--50~keV intensity reaches the maximum
before the hard-to-soft transition, and the other type does 
after the transition.
There is a difference in 
the duration of
the initial hard state, from the outburst onset to
the hard-to-soft transition.
The durations of seven outbursts were longer than 
\textcolor{black}{
$\sim$9~d,
}
while the other three ones were shorter than that.
Therefore we named the former type as 
slow-type (S-type) and the latter type as fast-type (F-type).
These two types show the different properties
in the following four points:
(1) the duration from the outburst onset to the hard-to-soft transition
(``pre-transition time'') in the S-type is longer than that in the F-type;
(2) the 2--15~keV luminosity at the hard-to-soft state transition
(``transition luminosity'') in the S-type is larger than that in the F-type;
(3) the 15--50~keV luminosity reaches its maximum during the hard state
in the S-type but during the soft state in the F-type; and
(4) the luminosity before the outburst onset (``pre-outburst luminosity'')
in the S-type is larger than in the F-type.
Figure \ref{fig8} illustrates the schematic
light curves of the S-type and the F-type rising behaviors.
We here discuss how and why two types of the outburst rising are
created and evolved.

The difference between the S-type and the F-type begins with the
pre-outburst luminosity.
We consider the irradiation on the accretion disk during
the pre-outburst period as a key factor to separate the
S-type from the F-type.
The hard-to-soft state transition is
generally considered as the transition of the accretion disk according
to the change of the accretion rate (e.g., \cite{Mineshige1989};
\cite{Abramowicz1995}). 
The X-ray irradiation on the disk from inner
regions can affect the disk structure (\cite{Kim1999};
\cite{Dubus2001}, and referenced therein).
If mass accretion from the companion star increases,
the accretion disk is usually cooled down by
the increased thermal soft X-ray emission.
As the result, the geometrically thick disk turns to the thin disk,
which corresponds to the hard-to-soft state
transition.
If an intense disk irradiation exists, the thick-to-thin
disk transition is suppressed because the irradiation heats up the
disk and keep the geometrically thick structure.
Therefore, the pre-transition time becomes longer.
This agrees with the observed feature that the pre-transition time
in the S-type is longer than that in the F-type. 

The source of the irradiation emission has several candidates such as
blackbody emissions from the surface of the neutron-star and/or
Comptonized emissions from the corona somewhere around the disk. 
The long pre-transition time is considered to cause the large ``transition
luminosity'' in the S-type
\textcolor{black}{
The present results show that the S-type transition tends to occur
at higher luminosity than the F-type transition does.
Most of the S-type transitions occurred at the higher
than 4\% of the Eddington luminosity for 1.4\Mo.
}
The different transition luminosity between the S-type and the F-type
reminds us of a hysteretic behaviour in the state transition that has been
reported in both BH-LMXBs and NS-LMXBs (\cite{Miyamoto1995};
\cite{MacCop2003}; Yu et al.2004).
We thus consider that these root causes have some relation.
\citet{Yu2004} proposed a
``two-accretion-flow geometry'' to explain the hysteresis.
According to their model, in case of the S-type, a sub-Keplerian flow (i.e.,
corona) may be initially irradiated by the pre-outburst radiation, and
then the disk would be heated up to prevent a transition.  To explain
this behavior furthermore, theoretical calculations considering
magneto-hydro dynamics are required, similar to those for BH-LMXB
disks introduced by \citet{Oda2010}.

In the S-type, the 15--50~keV luminosity gets the maximum during the
initial hard state, whereas it does after the hard-to-soft transition
in the F-type.
\textcolor{black}{
The difference can be explained by the different values
in the transition luminosity as follows.
Both the 2--15~keV and 15--50~keV luminosities change similarly
before/after the state transition.
The 15--50~keV luminosity drops at the hard-to-soft transition
in both the S-type and the F-type,
although the drop is small in the F-type.
}
In the S-type, the transition luminosity is
relatively bright and the 15--50~keV luminosity drops so greatly at
the transition that it will never recover to the level of the
pre-transition period during the soft state.
On the other hand, in the F-type,
the transition luminosity is relatively faint and the drop at the
transition is so small that both the 2--15~keV and the 15--50~keV
luminosities reach the maximum after the transition.

The present results also resemble the hard-to-soft transition
behavior in BH-LMXBs reported by \citet{Gierlinski2006}.
They obtained two types of transitions classified into the bright/slow or
the dark/fast in the initial part of their outbursts.
The bright/slow transition occurs at $\sim$30\% of the Eddington
luminosity and takes $\gtrsim$30~d during which the source quickly reaches
the intermediate/very high state and then proceeds to the soft state at
slower pace.
The dark/fart is less luminous ($\ltsim$10\% of the
Eddington luminosity), shorter ($\ltsim$15~d) period, and the source
does not slow its transition rate before reaching the soft state.
It is quite similar to our results of NS-LMXBs in a point that there
are two types of transitions in the initial part of the outbursts.

However, the parameters representing the transition duration are not
exactly the same.
We employed the ``pre-transition time'' from the
outburst onset to the hard-to-soft transition for this, whereas
\citet{Gierlinski2006}
defined it by the period from the last hard-state time to first
soft-state time and refereed it as the ``transition duration''.
As for NS-LMXBs, the ``transition duration'' is very short
($\ltsim$1~d) in
\textcolor{black}{most}
outbursts, and their difference cannot
be recognized from the daily light curve.
However, they are common in a point that the two types
are different in the transition luminosity.
\citet{Gierlinski2006}
speculated that the distinction of the two types is due to irradiation
and evaporation of the disk.  Hence, we note that the behavior of
outbursts of BH-LMXBs is similar to those of NS-LMXBs.

We have found two distinct groups of the fast and slow types in the
outburst initial behavior from the data of only two soft transients,
Aql~X-1 and 4U~1608$-$52.
It is a future issue whether the present
result is common to general soft X-ray transients containing a
neutron-star.  Therefore, we expect that one pays attention to
generalize the present result.

\bigskip
The authors would like to acknowledge the MAXI team for MAXI operation
and for watching and analyzing real time data.  They also thank the
RXTE-ASM team and the Swift-BAT team for providing excellent data
publicly.  This research was partially supported by the Ministry of
Education, Culture, Sports, Science and Technology (MEXT),
Grant-in-Aid for Science Research 20244015.

\appendix
\section*{Outburst Onset}

\begin{figure*}
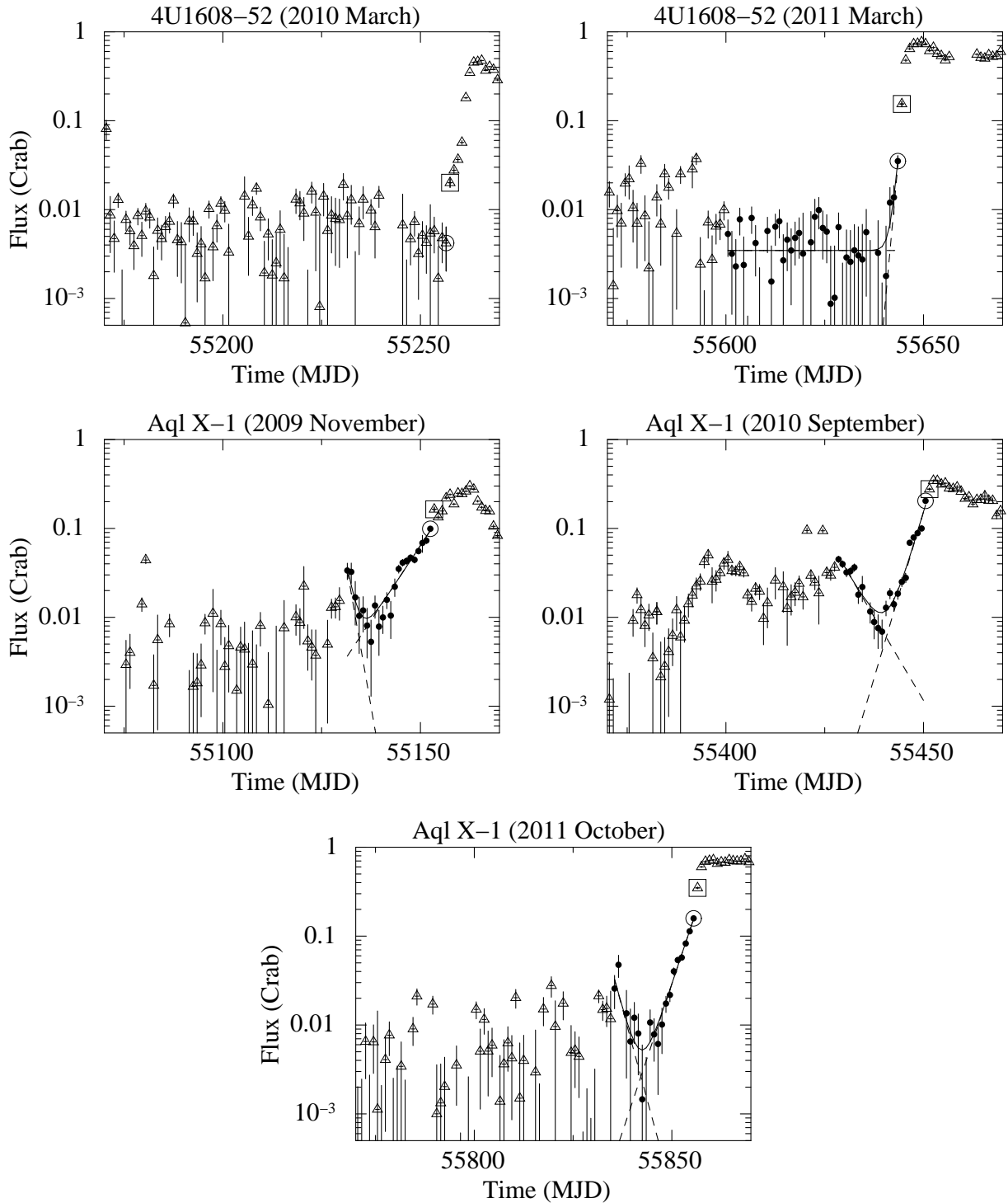

  \begin{center}
    \FigureFile(80mm,){fig9-a.eps}
    \FigureFile(80mm,){fig9-b.eps}
    \FigureFile(80mm,){fig9-c.eps}
    \FigureFile(80mm,){fig9-d.eps}
    \FigureFile(80mm,){fig9-e.eps}
  \end{center}
  \caption{MAXI-GSC light curves (2--20~keV) of five outbursts
around the onset fitted with two exponential functions.
The solid curve shows the best-fit model.
The dashed lines show each components. 
Filled circles represent data points used for the model fit.
Open circles indicate the data point of ``last hard state''.
Open squares indicate the data point of ``first soft state''.
}
\end{figure*}

\begin{figure*}
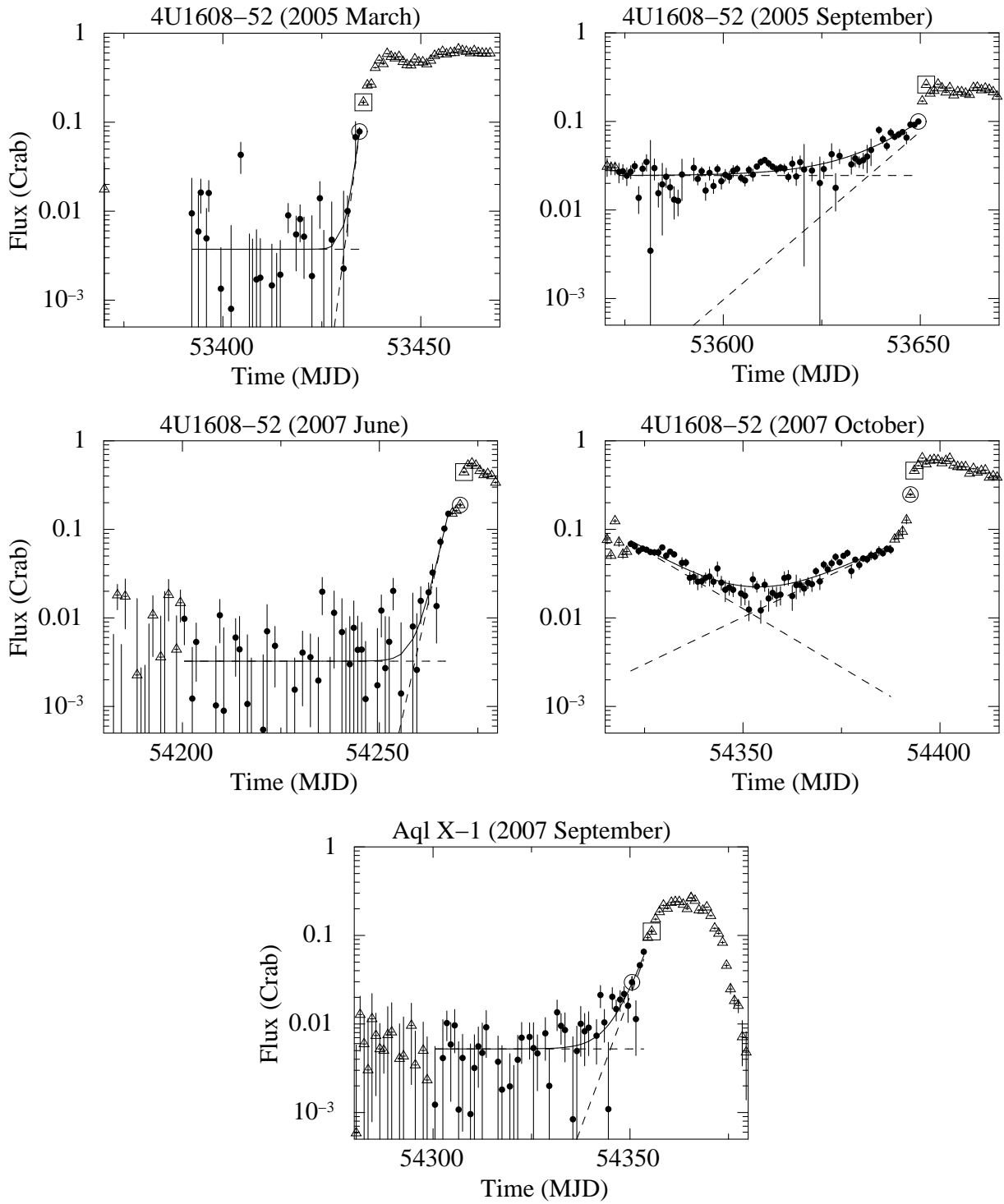

  \begin{center}
    \FigureFile(80mm,){fig10-a.eps}
    \FigureFile(80mm,){fig10-b.eps}
    \FigureFile(80mm,){fig10-c.eps}
    \FigureFile(80mm,){fig10-d.eps}
    \FigureFile(80mm,){fig10-e.eps}
  \end{center}
  \caption{RXTE-ASM light curves (2--10~keV) of five outbursts
around the onset fitted with two exponential functions.
Marks are the same as in figure~9.}
\end{figure*}

We define the outburst onset as the start point
of the increase towards the outburst peak.
We estimated the outburst onset from the GSC or ASM light curves
with a 1-day time bin.
We fitted the light curves around outburst onset with a model
of two exponential functions.
When the persistent component was constant within the error,
we used a model with one constant and one exponential function.
We determined the cross point of these two functions as the outburst onset.
The errors are at the 90~\% confidence level.
In the cases of the outbursts of 4U~1608$-$52 in 2007 October and 
Aql~X-1 in 2010 September, probably due to the variable persistent flux,
the best-fit models are not formally
acceptable.  
We estimated the errors in consideration of the fitting range. 
We explain how the outburst began in each outburst below.

\begin{description}
\item[4U~1608$-$52, 2010 March (figure~9a):]
The flux before outburst onset stayed in the almost
constant level. After MJD=55256.5 (last hard state),
the flux increased clearly, and next data point is ``first soft state''.
Thus, we determined the ``last hard state'' as the ``outburst onset''.
\item[4U~1608$-$52, 2011 March (figure~9b):]
The flux before outburst onset stayed in the almost constant level.
We fitted a model with one constant and one exponential function
between MJD=55600 and MJD=55644.
\item[Aql~X-1, 2009 November (figure~9c):]
The flux decreased between MJD$\sim$55130 and $\sim$55140.
After that, the flux increased towards the peak of the outburst.
We fitted a model with two exponential functions
between MJD=55131 and MJD=55153.
\item[Aql~X-1, 2010 September (figure~9d):]
The flux decreased between MJD$\sim$55248 and $\sim$55440.
After that, the flux increased towards the peak of the outburst.
We fitted a model with two exponential functions
between MJD=55428 and MJD=55451.
However, because the slope changed during the increase,
the fit is not acceptable.
We estimated the errors in consideration of the change of slope.
\item[Aql~X-1, 2011 October (figure~9e):]
The flux decreased between MJD$\sim$55835 and $\sim$55845.
After that, the flux increased towards the peak of the outburst.
We fitted a model with two exponential functions
between MJD=55835 and MJD=55856.

\item[4U~1608$-$52, 2005 March (figure~10a):]
The flux before outburst onset stayed in the almost constant level.
We fitted a model with one constant and one exponential function
between MJD=53390 and MJD=53435.
\item[4U~1608$-$52, 2005 September (figure~10b):]
The persistent trend  before outburst onset was considered to stay
in the almost constant level, although the flux is variable.
We fitted a model with one constant and one exponential function
between MJD=53573 and MJD=53650.
\item[4U~1608$-$52, 2007 June (figure~10c):]
The flux before outburst onset stayed in the almost constant level.
We fitted a model with one constant and one exponential function
between MJD=54200 and MJD=54268.
\item[4U~1608$-$52, 2007 October (figure~10d):]
The flux decreased between MJD$\sim$54320 and $\sim$54350.
After that, the flux increased towards peak of the outburst.
We fitted a model with two exponential functions
between MJD=54321 and MJD=54388
However, because the slope changed during the increase,
the fit is not acceptable.
We estimated the errors in consideration of the change of inclination.
\item[Aql~X-1, 2007 September (figure~10e):]
The flux before outburst onset stayed in the almost constant level.
We fitted a model with one constant and one exponential functions
between MJD=54300 and MJD=54354.
\end{description}

\end{document}